\documentclass[a4paper,11pt]{article}
\pdfoutput=1 

\usepackage{jheppub} 
\usepackage{color}
\usepackage[dvipsnames]{xcolor}
\usepackage{newlfont}
\usepackage{graphicx}
\usepackage{amssymb}
\usepackage{amsmath}
\usepackage{float}
\usepackage{amsfonts}
\usepackage{tablefootnote}
\usepackage{bm}
\usepackage{verbatim}
\usepackage[latin1]{inputenc}
\usepackage{bbm}
\usepackage{multirow}
\usepackage[thinlines]{easytable}
\usepackage{braket}
\usepackage{amsthm}
\usepackage{bbold}
\usepackage{subfig}
\usepackage{mathrsfs}
\usepackage[varg]{txfonts} 
\usepackage{amsfonts}

\newcommand{\ba}{\begin{array}}
	\newcommand{\ea}{\end{array}}

\DeclareMathOperator{\Tr}{Tr}

\newcommand{\mck}{\mathcal{K}}

\newcommand{\mcc}{\mathcal{C}}
\newcommand{\tbfr}{\textbf{r}}
\newcommand{\red}{\textcolor{red}}

\newcommand{\TFD}{\text{TFD}}
\newcommand{\orange}{\textcolor{black}}

\hypersetup{linkcolor=red,          % color of internal links
	citecolor=teal,        % color of links to bibliography
	urlcolor=blue}          % color of the URL

\title{Higher-Order Krylov State Complexity in Random Matrix Quenches}

  \author[a]{Hugo A. Camargo,}
  \author[a]{Yichao Fu,}
  \author[a]{Viktor Jahnke,}
  \author[a,b]{Keun-Young Kim,}
  \author[a]{Kuntal Pal}
 \affiliation[a]{
	Department of Physics and Photon Science, Gwangju Institute of Science and Technology,
    \\123 Cheomdan-gwagiro, Gwangju 61005, Republic of Korea}
 \affiliation[b]{Research Center for Photon Science Technology, Gwangju Institute of Science and Technology, 
 \\123 Cheomdan-gwagiro, Gwangju 61005, Korea}

\emailAdd{hugo.camargo@gist.ac.kr, yichao.fu@gm.gist.ac.kr, viktorjahnke@gist.ac.kr, fortoe@gist.ac.kr, kuntalpal@gist.ac.kr }

\abstract{ 
In quantum many-body systems, time-evolved states typically remain confined to a smaller region of the Hilbert space known as the {\it Krylov subspace}. The time evolution can be mapped onto a one-dimensional problem of a particle moving on a chain, where the average position $\langle n \rangle$ defines Krylov state complexity or spread complexity. Generalized spread complexities, associated with higher-order moments $\langle n^p \rangle$ for $p>1$, provide finer insights into the dynamics.
We investigate the time evolution of generalized spread complexities following a quantum quench in random matrix theory. The quench is implemented by transitioning from an initial random Hamiltonian to a post-quench Hamiltonian obtained by dividing it into four blocks and flipping the sign of the off-diagonal blocks. This setup captures universal features of chaotic quantum quenches.
When the initial state is the thermofield double state of the post-quench Hamiltonian, a peak in spread complexity preceding equilibration signals level repulsion, a hallmark of quantum chaos. We examine the robustness of this peak for other initial states, such as the ground state or the thermofield double state of the pre-quench Hamiltonian. To quantify this behavior, we introduce a measure based on the peak height relative to the late-time saturation value. In the continuous limit, higher-order complexities show increased sensitivity to the peak, supported by numerical simulations for finite-size random matrices.
}

\begin{document}
\maketitle
	
%%%%%%%%%%%%%%%%%%%%%%%
%%%%%%%%%%%%%%%%%%%%%
\section{Introduction}

Recently, notions of complexity based on Krylov subspace methods have gained significant attention\footnote{Krylov complexity for operators was first introduced in \cite{Parker:2018yvk}, while its extension to states was introduced in \cite{Balasubramanian:2022tpr}. See \cite{Nandy:2024htc} for a concise review of different aspects of quantum dynamics in the Krylov space.}. This growing interest stems from two key factors. First, emerging evidence suggests a connection between Krylov complexity and holographic complexity~\cite{Rabinovici:2023yex}, which describes the growth of the black hole interior in the AdS/CFT duality. Studying Krylov complexity within holographic systems may therefore, provide new insights into black holes and quantum gravity. Second, Krylov complexity -- whether applied to quantum states or operators -- provides a novel framework for characterizing quantum chaotic behavior, potentially bridging the gap between different notions of quantum chaos. This is particularly important, as quantum chaos plays a crucial role in various domains, including understanding thermalization in closed quantum systems \cite{DAlessio:2015qtq}, characterizing the scrambling properties of black holes (cf. the review \cite{Jahnke:2018off}), and demonstrating quantum advantage in quantum computing \cite{Arute2019}.

There are several approaches to characterizing quantum chaotic behavior. Two of the most widely used methods are out-of-time-order correlators (OTOCs) and spectral statistics of the Hamiltonian. OTOCs measure the scrambling of quantum information and are typically used to study quantum chaos at early times, up until the scrambling time $t_*$. In chaotic systems with finite Hilbert space dimension, OTOCs are expected to reach an equilibrium value close to zero for $t \gtrsim t_*$ for almost any choice of operators. Spectral statistics, on the other hand, are associated with longer timescales. Key metrics in this framework include nearest-neighbor level spacing statistics, which quantify level repulsion, and the spectral form factor, where a characteristic `ramp' indicates spectral rigidity. The spectral statistics of chaotic systems are expected to align with those predicted by random matrix theories of the appropriate ensemble. The notion of Krylov complexity allows us to probe both early-time and late-time quantum chaos within a unified formalism, with Krylov operator complexity corresponding to early-time probes of quantum chaos, while Krylov state complexity (commonly also referred to as the spread complexity) relates to late-time probes\footnote{In systems with finite Hilbert space dimension, one can also use the saturation value of Krylov operator complexity to diagnose chaos at late times~\cite{Rabinovici:2022beu}.}. The basic idea behind Krylov complexity is that, under time evolution, both operators and states do not explore the full Hilbert space, but rather a fraction of it, called the Krylov subspace. By constructing an orthonormal basis for this subspace, known as the Krylov basis, the time evolution of the operator or state can be mapped onto a one-dimensional problem of a particle moving on a chain, where higher positions on the chain correspond to more complex operators or states. The motion of the particle along the chain is described by hopping terms, $b_n$ and $a_n$, known as the Lanczos coefficients.  Krylov operator and state complexities in this scenario are then defined as the average position of the particle on the chain, denoted as $\langle n \rangle$, and their time dependence typically reveals whether the system's dynamics are chaotic or integrable\footnote{Similarly to OTOCs, Krylov complexity fails to correctly identify the dynamics as chaotic or integrable in the presence of saddle dominating scrambling~\cite{Bhattacharjee:2022vlt, Huh:2023jxt}. Additionally, Krylov operator complexity is not a good probe for chaos in quantum field theories with continuous spectrum, since in this case the analytic properties of the Wightman two-point function imply that, at a finite temperature $T$,  $\lambda_K =2\pi T$, regardless of whether the theory is free or interacting. See for instance~\cite{Dymarsky:2021bjq, Avdoshkin:2022xuw, Camargo:2022rnt, Kundu:2023hbk}. }. %\HC{In the foodnote here we use $\beta$ as the inverse temperature, but in the rest of the draft we use it to signify Dyson index. Maybe here we can write $\lambda_K =2\pi T$ instead? } \red{ Good point. I have corrected it, hopefully Viktor does not mind it. Anyway foodnote sounds fun!} \HC{XD. haha you may erase my comment.}

{\bf Krylov Operator complexity} quantifies how local operators spread over time within the Krylov subspace. In chaotic systems at the thermodynamic limit and in a thermal state, the Lanczos coefficients $b_n$ exhibit unbounded linear growth (with logarithmic corrections for one-dimensional systems), leading to an exponential growth of Krylov operator complexity after an initial quadratic phase. This growth reflects how rapidly the operator expands into more complex directions in the Hilbert space. Integrable or weakly chaotic systems, however, show a slower, typically non-exponential, growth, indicating that the operator's complexity increases more gradually. Interestingly, the rate of this exponential growth, denoted by $\lambda_K$, is conjectured to establish an upper bound on the Lyapunov exponent $\lambda_L$, which governs the behavior of out-of-time-order correlators (OTOCs), such that $\lambda_L \leq \lambda_K$~\cite{Parker:2018yvk}\footnote{This bound is rigorously proven at infinite temperature; see ~\cite{Parker:2018yvk}.}. Away from the thermodynamic limit, as the system explores only a finite-dimensional Krylov subspace, the exponential growth of Krylov operator complexity is expected to persist up to a time scale given by $t\sim \log N$, where $N$ is proportional to the number of degrees of freedom. After that Krylov operator complexity grows linearly until saturation at $t\sim e^{2N}$~\cite{Rabinovici:2022beu}. In chaotic systems within fixed symmetry sectors, Krylov operator complexity saturates at a value around $K/2$, where $K$ is the dimension of the Krylov space, whereas for integrable systems it saturates at lower values~\cite{Rabinovici:2022beu}.

{\bf Krylov state complexity or Spread Complexity} measures the spread of a quantum state within the Krylov subspace over time. Starting the time evolution from an initial state, such as a thermofield double state (TFD), the complexity initially grows quadratically with time in chaotic systems, followed by linear growth, and then eventually reaches a peak before plateauing. This growth pattern -- quadratic, linear, peak, and plateau -- mirrors the behavior seen in the spectral form factor for systems whose spectral statistics match those of random matrix theory, including random matrix models and Hamiltonian systems following random matrix theory statistics~\cite{Balasubramanian:2022dnj, Erdmenger:2023wjg}. Similar growth behavior has been observed in models such as spin chains~\cite{Camargo:2024deu,Alishahiha:2024vbf, Gautam:2023bcm}, the SYK model~\cite{Baggioli:2024wbz, Balasubramanian:2022tpr, Erdmenger:2023wjg}, and chaotic billiards~\cite{Hashimoto:2023swv, Balasubramanian:2024ghv}. In contrast, integrable or weakly chaotic systems do not exhibit a peak; instead, the complexity saturates more gradually, reaching lower final values~\cite{Alishahiha:2024vbf, Balasubramanian:2022tpr, Baggioli:2024wbz, Huh:2024lcm}. Both in chaotic and integrable systems, the saturation timescale is given by the system's Hilbert space dimension. \footnote{We refer to refs. \cite{spread1,spread2, Gill:2024acg, Bhattacharya:2023yec, Bhattacharya:2024hto, Du:2022ocp, Iizuka:2023fba, Bento:2023bjn, Loc:2024oen, Malvimat:2024vhr, Bhattacharyya:2023grv,  Gautam:2023pny, He:2024xjp, Li:2024ljz, Ganguli:2024myj, Bhattacharjee:2024yxj, Jeong:2024oao, Caputa:2024vrn, Camargo:2023eev, Fan:2023ohh} for various other recent interesting applications of Krylov and spread complexity.}

Since the average position of the particle on the Krylov chain characterizes the dynamics of the system, it is natural to study higher moments, for example, $\langle n^2 \rangle$ or $\langle n^3 \rangle$, and thus obtain a more complete characterization of how the particle moves in the chain. One can define the so-called {\it generalized spread complexities}, $C_p = \langle n^p \rangle$, where $p=1,2,3,...$, which can be shown to be related to the probability distribution associated with measurements of the spreading operator~\cite{Kuntal-YiChao}. Similar to the spread complexity, the generalized spread complexities $\langle n^p \rangle$ are also expected to display a rise-peak-plateau structure in chaotic systems, as suggested by the results obtained in random matrix theory using a continuous approximation~\cite{Kuntal-YiChao}. This result, however, assumes that the initial state is a random initial state or an infinite temperature thermofield double state. It is clear, however, that the above-mentioned results, especially the presence of the peak before saturation, are state-dependent. For example, suppose one considers an eigenstate of the Hamiltonian. In that case, the time evolution is trivial, given by a phase, and the time-evolved state does not overlap with other states in the Hilbert space. Correspondingly, the generalized spread complexity is identically zero. That naturally raises the question of whether there are other states for which the generalized spread complexity displays a peak after saturation. What are the conditions on the initial state and on the system for the peak in spread complexity to be present, revealing details about the dynamics? A very natural setup in which we can study the state-dependence of generalized spread complexity and answer the above question is provided by {\it quantum quenches}. 

Quantum quenches provide a framework for investigating the nonequilibrium dynamics of closed, interacting quantum systems following a change in one or more of the system's parameters. Initially, the system is prepared in a quantum state $ |\psi_0 \rangle $, which is usually the ground state or a linear combination of the eigenstates of the pre-quench Hamiltonian $H_0$. A modification to one of the system's parameters transforms $H_0$ into the post-quench Hamiltonian $H$. The focus is then on the time evolution of $|\psi_0 \rangle$  under the dynamics generated by the new Hamiltonian $H$, providing insights into how the system relaxes and evolves within its new configuration. If the dynamics are chaotic, the system is expected to evolve toward an equilibrium state as predicted by statistical mechanics. The characteristic time scale of equilibration depends on the observable in question. Generalized spread complexity, for instance, is expected to thermalize only at very long times, on the Heisenberg time scale, which is proportional to the dimension of the system's Hilbert space. 

In this work, we investigate the time evolution of generalized spread complexity following quantum quenches within the framework of random matrix theory, whose universal properties are expected to capture a wide range of chaotic systems. Building on the protocol proposed in \cite{Brandino}, we consider quantum quenches based on random matrices sampled from the Gaussian orthogonal ensemble (GOE) and extend these results to matrices drawn from the Gaussian unitary ensemble (GUE). We expect the GOE-based quench to model quenches in chaotic systems with time-reversal symmetry, while GUE-based quenches are anticipated to describe systems without time-reversal symmetry, such as those in the presence of a magnetic field.

The basic idea is to sample the initial pre-quench Hamiltonian \( H_0 \) from an ensemble of \( N \times N \) random matrices with even \( N \). We then split the matrix into four equal blocks: two on the diagonal and two off-diagonal. The post-quench Hamiltonian is obtained by flipping the sign of all elements in the off-diagonal blocks. This setup allows us to systematically investigate the time-dependence of generalized spread complexity for various initial states, including the ground state and the thermofield double state of the pre-quench Hamiltonian, both evolved with the post-quench Hamiltonian. We also compare these results to those obtained when the initial state is the thermofield double state of the post-quench Hamiltonian, which is the state commonly considered in the literature. In particular, we explore how the peak in the time-dependence of generalized spread complexities \( C_1, C_2, C_3, \) and \( C_4 \) depends on the choice of initial state. We propose a method to precisely quantify the peak's height by comparing it with the late-saturation value of spread complexity. In the continuous limit, we show that higher-order generalized complexities are more sensitive to the presence of the peak, and we confirm this expectation through numerical simulations.

This paper is structured as follows. 

\begin{itemize}
\item Section 2 - We briefly discuss the motivation behind introducing the generalized spread complexity and study its time evolution behavior in the case of the Hamiltonian element of the GOE. To quantify the characteristics of the peak in the time evolution of complexity for a chaotic Hamiltonian, we define a parameter whose positive value signals the presence of the peak, while its value signifies the sharpness of the peak. We also obtain the numerical values of these parameter from the 
analytical expressions of the generalized spread complexity in the continuum limit. 

\item Section 3 - In this section, we study the spreading of an initial wavefunction after a non-adiabatic sudden quench by modeling the system Hamiltonian through a one-parameter family of random matrices which, for the special values of the parameter, can belong to the GOE (or GUE). The time evolution of several orders of the generalized spread complexity 
is studied in detail for different initial states before the quench. How different qualitative 
features of the initial state affect the corresponding generalized spread complexity is discussed by studying the inverse participation ratio and the distribution of the ratio of successive Lanczos coefficients. 

\item Section 4 - We provide a summary of the main results of the paper and discuss some of their relevant implications. 

\item This paper also contains three appendices, where we discuss (A) an analytical model of a sudden quench between two harmonic oscillators; (B) properties of Gaussian $\beta$-ensembles and the distribution of the ratio of two successive Lanczos coefficients; (C) the relation between the quench parameter of the random matrix Hamiltonian and the $r$-parameter of the energy spectrum. 
\end{itemize}

%%%%%%%%%%%%%%%%%%%
%%%%%%%%%%%%%%%%%
 \section{Spread complexity and its higher order generalizations} \label{definitionGSC}
Consider the time evolution of an initial state $\ket{\psi_0}$ generated by a time-independent Hamiltonian $H$. 
The primary quantity we study in this paper is the so-called generalized spread complexity defined as
\begin{equation}\label{GSC_definition}
    \mathcal{C}_m(t)=\sum_n n^m ~|\braket{\Psi(t)|K_n}|^2 = \sum_n n^m ~|\phi_n(t)|^2 ~,
\end{equation}
where the state $\ket{\Psi(t)}$ is the initial state evolved in time by the Hamiltonian $\ket{\Psi(t)}=e^{-itH} \ket{\psi_0}$ and $\ket{K_n}$ are the elements of the Krylov basis generated by the Hamiltonian $H$ starting from the initial state $\ket{\psi_0}$ at the start of the evolution at $t=0$. Here, $m$ are integers, $m=1,2,3\cdots$, and when $m=1$, one gets back the definition of the spread complexity originally introduced in \cite{Balasubramanian:2022tpr}. 

It is useful to introduce the following operator which is diagonal in the Krylov basis, 
and is known as the generalized spreading operator,
\begin{equation}
		\mck_m =  \sum_n  n^m \ket{K_n} \bra{K_n}~.
\end{equation}
Using this definition, it is easy to see that generalized spread complexity in eq. \eqref{GSC_definition} is just the expectation value of the generalized spreading operator in the time-evolved state,\footnote{In the context of operator growth, the generalized Krylov complexity was studied in \cite{Fan:2023ohh}.}
\begin{equation}
		\mathcal{C}_m(t)=  \bra{\Psi(t)} \mck_m \ket{\Psi(t)}~=  \bra{\psi_0} \mck_m (t) \ket{\psi_0}~,~~~\text{with}~~\mck_m(t)= e^{i H t} \mck_m e^{-iHt}~.
\end{equation}
 
A convenient way to understand the origin of generalized spread complexity -- and our motivation for exploring its generalization to higher orders of $m$ as introduced in \cite{Balasubramanian:2022tpr} -- is to consider the following generating function:
\begin{equation}\label{generating}
		G(\eta,t)= \sum_{n} e^{\eta n}  |\phi_n(t)|^2 ~,
\end{equation}
where $\eta$ is an auxiliary parameter.
A Taylor series expansion of this quantity around $\eta =0$ shows that the coefficients of this expansion
are precisely equal to $ \mathcal{C}_m(t)$. Therefore, one can analytically continue $\eta$ 
to complex values and obtain the characteristics function $\chi_{\mck} (u,t)$ of the 
distribution of the spreading operator $\mck=\mck_1$, i.e., \cite{Kuntal-YiChao}
\begin{equation}\label{CFSC}
		\chi_{\mck} (u,t) = G(-iu,t)=\bra{\psi_0} e^{-iu \mck(t)} \ket{\psi_0}~.
\end{equation}
All higher-order moments of the spreading operator then naturally arise from this characteristic function and define the corresponding higher-order generalized spread complexities. 

Before moving on to consider the time evolution of different orders of $\mcc_m(t)$, we discuss an interpretation of these
quantities in terms of a hopping problem of a fictitious particle in the discrete Krylov chain. In the approach of studying the quantum dynamics in the Krylov subspace, one essentially maps
any unitary time evolution to a hopping problem in a one-dimensional discrete lattice, commonly called the Krylov chain. Then, the spread complexity $\mcc_1(t)$ represents the average 
position of a factitious particle in this chain, while higher order $\mcc_m(t)$ provides 
information about the higher order moments of the particle's position. 
To see this more clearly, we note that the Fourier transformation of the characteristics function $\chi_{\mck} (u,t)$ is the distribution of the result of a measurement of the location of this particle in the chain at any arbitrary time $t$, i.e., 
\begin{equation}\label{SOsta}
		P(j, t)  = \int e^{iu O} ~ \chi_{\mck} (u,t)~ du~= \sum_n | \bra{ K_n} e^{-it H} \ket{\psi_0}| ^2 \delta (j-n)~=  \sum_n   |\phi_n(t)|^2 ~ \delta (j-n)~.
\end{equation}
The spread complexity is the mean, while the higher-order complexities are higher-order moments of this distribution. 

Each of these quantities (which are computed in the Krylov basis) minimizes a corresponding cost function defined with respect to an arbitrary complete orthonormal basis in the Hilbert space for a finite time for continuous time evolution \cite{Balasubramanian:2022tpr}. However, we also note that any arbitrary linear combination of these quantities may not be the minimum in the Krylov basis. 
In that sense, these combinations are not `good' measures of the complexity of wavefunction 
spreading. E.g., the variance of the position of the fictitious particle in the Krylov chain, 
$(\Delta n(t))^2 = \braket{n^2} - \braket{n}^2~=\mcc_2(t)-\mcc_1(t)^2$ is not a measure of complexity, and for this reason, we do not study this quantity further in this paper.

It is well known that the spread complexity $\mcc_1(t)$ grows quadratically at early times, irrespective of the initial state or the system Hamiltonian \cite{Balasubramanian:2022tpr,Erdmenger:2023wjg}. 
One can similarly show that at early times $\mcc_m(t)$ of all orders also show the same quadratic growth. To see this we start from values of the first few derivatives of $\phi_n(t)$ at early times, which up to second order derivative are given by, $\phi_n(0)=\delta_{n0}$, $\dot{\phi}_n(0)= -i (a_0 \delta_{n0} + b_1 \delta_{n1}) $, and $\ddot{\phi}_n(0)= -(a_0^2+b_1^2) \delta_{n0}- a_0 b_1 \delta_{n1}-b_1 b_2 \delta_{n2}$. Now Taylor expanding the definition of $\mcc_m(t)$ in eq. \eqref{GSC_definition} up to second order, we see that it is given by $\mcc_m(t) \approx b_1^2 t^2 \sum_n n^m \delta_{n1} +\mathcal{O}(t^3) $. Therefore, $\mcc_m(t)$s of all orders show quadratic growth at very early times after the start of the evolution. 

We now discuss some typical properties of the generalized spread complexities under time evolution by chaotic quantum many-body systems, which will be useful later in this paper.  Consider a Hamiltonian $H$ with Hilbert space $\mathcal{H}$.
Following~\cite{Balasubramanian:2022tpr}, we consider the tensor product of the Hilbert space with itself $\mathcal{H}_L \otimes \mathcal{H}_R$. The initial state is taken to be the infinite temperature thermofield double (TFD) state defined as  
\footnote{Since this paper focuses exclusively on the infinite-temperature case, we shall refer to this state simply as the TFD state, without explicitly specifying that it corresponds to the infinite-temperature limit of the finite-temperature TFD state. For notational simplicity, we denote this state as $|\TFD\rangle$ throughout the manuscript, omitting the $\infty$ symbol in the subscript.}
\begin{equation}\label{TFD_post}
    \ket{\TFD}_{\infty}=\frac{1}{\sqrt{N}} \sum_n \ket{n} \otimes \ket{n}~,
\end{equation}
where $\ket{n}$ denotes the eigenstates of the Hamiltonian $H$. This state is invariant under time evolution with the Hamiltonian of the form $H_L \otimes \mathbb{I}-\mathbb{I} \otimes H_R $, where $H_L=H_R=H$, but is not invariant under time evolution generated by a single Hamiltonian. We choose to study the time evolution of the initial state with a Hamiltonian of the form $H_L \otimes \mathbb{I}$ \footnote{Note that this is equivalent to considering a single copy of the Hilbert space and choosing a maximal superposition state, $\ket{\psi} = \frac{1}{\sqrt{N}} \sum_n \ket{n}$, as the initial state. The perspective of using a double copy of the Hilbert space is convenient for interpreting the spectral form factor as the survival amplitude of the TFD state. While this perspective is not essential to our work, we adopt it to align with the nomenclature used in previous studies on this topic~\cite{Balasubramanian:2022tpr, Erdmenger:2023wjg}}.

\begin{figure}[h!]
		\centering
		\includegraphics[width=3in,height=2.2in]{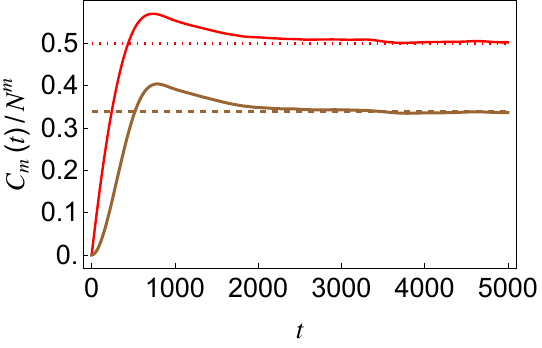}
		\caption{Evolution of $\mathcal{C}_1(t)$ (red) and $\mathcal{C}_2(t)$ (brown) for a random matrix Hamiltonian belonging to the GOE. Here, the initial state is the infinite temperature TFD state constructed from the eigenstates of the Hamiltonian with $N=1000$, and we have averaged over ten different realizations of the GOE matrix. The red and brown dashed lines indicate the saturation values of $\mathcal{C}_1(t)/N$ and $\mathcal{C}_2(t)/N^2$, which for the infinite temperature TFD state are, respectively,  1/2 and 1/3.}
		\label{fig:c1c2goe1000}
	\end{figure}

After drawing the Hamiltonian \(H\) from a random matrix ensemble (either GOE or GUE), we apply the Lanczos algorithm and compute the generalized spread complexity of different orders. In Fig. \ref{fig:c1c2goe1000}, we plot the evolution of the spread complexity, \(\mcc_1(t)\), and the second-order generalized spread complexity, \(\mcc_2(t)\), for a Hamiltonian belonging to the GOE with dimension \(N=1000\), using the TFD state as the initial state.
From the plots, it can be seen that while the general time evolution of both \(\mathcal{C}_1(t)\) and \(\mathcal{C}_2(t)\) follows the expected pattern of rise, peak, slope, and plateau for a chaotic Hamiltonian, there are certain differences between them. At early times, both exhibit quadratic growth (which is consistent with the discussion above). However, for the spread complexity \(\mcc_1(t)\), this growth quickly transitions to a linear regime, while for \(\mcc_2(t)\), the quadratic growth persists for a longer period before merging into a linear growth.
Additionally, the peak in \(\mcc_2(t)\) occurs later than in \(\mcc_1(t)\). This behavior holds generally for higher-order generalized spread complexities: as \(m\) increases, the peak occurs progressively later. Furthermore, the height of the peak (measured from the saturation value) in \(\mcc_2(t)\) is greater compared to \(\mcc_1(t)\), and this property holds for all higher-order \(\mcc_m\)'s. As we will show in later sections, these characteristics of the generalized spread complexities hold for any generic initial state, not just for \(\ket{\mathrm{TFD}}\). Finally, when we consider the normalized quantities \(C_m/N^m \), the saturation value of the normalized spread complexity is higher than that of the normalized second-order generalized spread complexity. Note that, since the dynamics is chaotic, the saturation values correspond to the infinite-time averages of these quantities.  The saturation value of \(\mcc_m\) is primarily determined by the absolute value of the overlaps \(c_n = \braket{n|\psi_0}\) of the initial state $\ket{\psi_0}$ with the eigenstates \(\ket{n}\) as well as the diagonal elements of the spreading operator in the Krylov basis. 

%\red{In this context, we note that it is possible to obtain an upper bound on the saturation value of \(\mcc_m\) in terms of the inverse participation ratio (IPR) of the eigenstates in the initial state (see Appendix \ref{bound}).}

%%%%%%%%%%%%%%
%%%%%%%%%%%%%%
%%%%%%%%%%%%%
%%%%%%%%%%%%%

One way to quantify the peak in $\mcc_m(t)$ is to compare its magnitude with respect to the late time saturation value. To this end, we define the following parameter 
\begin{equation}\label{peak_parameter}
    P_m=\frac{C_m(t_{\text{peak}})-\bar{\mathcal{C}}_{m}}{C_m(t_{\text{peak}})}~.
\end{equation}
Here $t_{\text{peak}}$ stands for the time when $\mcc_m(t)$ reaches its peak value and 
$\bar{\mathcal{C}}_{m}$ is the infinite-time average of $\mcc_m(t)$.  We can find the 
time $t_{\text{peak}}$ by taking the derivative of $\mcc_m(t)$ and finding the lowest value of time when the derivative vanishes. When quantified in terms of the parameter $P_m$, the conclusion that higher order generalized spread complexity has a sharper peak is based on the fact that $P_{m_1}< P_{m_2}$ when $m_{1}<m_{2}$. 

One particular situation where it is possible to determine the parameter $P_m$ analytically for a random matrix Hamiltonian is in the so-called continuum limit. We discuss the procedure of obtaining $P_{m}$ for the first few values of $m$ in such a limit for a Hamiltonian belonging to GUE in the following subsection.

%%%%%%%%%%%%%%%%%%%%%%%%%
\subsection{Quantifying the peak in the generalized spread complexity in the  continuum limit}\label{peak_continuum}

In this subsection we first briefly discuss the continuum limit of the generalized spread complexity, and then use it to determine analytically the location of the peak and the parameter quantifying the height of the peak in $\mathcal{C}_m(t)$. 
In the continuum limit, one essentially assumes that the discrete Krylov basis index $n$ can be mapped to a continuous coordinate, $x_n=\epsilon n$, where $\epsilon$ is a small parameter, and the essential ingredients of the Krylov construction, namely the Lanczos coefficients $a_n$, $b_n$ and the functions 
$\phi_n(t)$ are smooth well-behaved functions of $n$, and hence that of $x_n$. This is a very useful approximation that one can make to solve for the wavefunctions in the Krylov basis, since, in the so-called first-order formalism, these
wavefunctions satisfy a first-order linear partial differential equation, which 
then can be solved to obtain the wavefunctions $\phi(t,x_n)$, and hence the 
generalized spread complexity in this limit for a given profile of the Lanczos coefficients 
\cite{Muck:2022xfc,Erdmenger:2023wjg}.  

The profile for the Lanczos coefficient depends on the initial state before the start of the 
evolution. Since, in this paper, we consider time evolution generated by a Hamiltonian  belonging to some Gaussian random matrix ensemble, we consider 
initial states for which 
the mean of the Lanczos spectrum can be approximated by the expressions in \eqref{LC_average}\footnote{
Usual examples of such initial states are a random state of the form $(1,0,0, \cdots,0)^T$  written in an appropriate basis, as well as states related to this by appropriate transformations depending on the ensemble under consideration.},
\begin{equation}\label{LC_average1}
    \braket{a_n}=0~,~~~\text{and}~~~ \braket{b_n}= \sqrt{1-\frac{n}{N}}~. 
\end{equation}
In this context it was pointed out  in \cite{Erdmenger:2023wjg}  that one can use the expressions in \eqref{LC_average1}
as an approximation for Lanczos coefficients associated with the maximally entangled infinite temperature thermofield double state constructed from the eigenstates of the 
Hamiltonian $H$ (and defined in a double-copy Hilbert space) in the following sense. If one uses a tridiagonal form for the Hamiltonian constructed from \eqref{LC_average1}, (i.e., uses zero as the diagonal elements and $\sqrt{1-n/N}$ as the subdiagonal elements, as $b_n$s), and apply the Lanczos algorithm with the maximally entangled state as the initial state then one gets back the same values as the Lanczos coefficients. Therefore, one can use the 
infinite temperature thermofield double state as the initial state, along with the 
profile for the Lanczos coefficients in \eqref{LC_average1} and proceed to analyses the 
generalized spread complexity in the continuum limit. 

Performing such an analysis for a random matrix Hamiltonian ($H$) 
belonging to the GUE ensemble, it is possible to obtain the analytical expressions for $\mcc_m(t)$ for different time intervals, and 
show that in the continuum limit, the ensemble-averaged generalized spread complexities have the 
characteristic rise-peak-slope behavior with time.  

The peak in the time evolution of $\mcc_m(t)$ appears at a time
$t<D$, where $D$ is the dimension of the Krylov subspace, which for the thermofield double state under consideration is the same as the rank of the random matrix $N$. Therefore to  determine 
the parameter $P$ it is sufficient to record the  expressions of $\mcc_m(t)$ during this time interval. The expressions for the normalized \footnote{The normalization is performed here
by dividing the expressions for ensemble-averaged $\mcc_m(t)$ by a suitable 
probability function and a factor of $N^m$. See \cite{Erdmenger:2023wjg} and \cite{Kuntal-YiChao} for details of this procedure.} 
ensemble-averaged spread complexity in terms of dimensionless time $v=t/N$, for $v<1$ is given by \cite{Erdmenger:2023wjg}
\begin{equation}
    \braket{\mathcal{C}^N_{1}(v)}= 
			\frac{7 \pi+640v -960 v^2+(320+20 \pi) v^3- 15 \pi v^4+3 \pi v^5}{2 (160+5 \pi -160 v + 15 \pi v^2 - 5 \pi v^3)}~,	~~\text{for}~~~ v<1~,
\end{equation}
while the expression for normalized ensemble-averaged $\mcc_2(t)$ in this time interval is given by  \cite{Kuntal-YiChao}
\begin{equation}
    \braket{\mathcal{C}^N_{2}(v)}=
			\frac{(19\pi/7)+640v^2-1280v^3+(800+10 \pi) v^4-(160+12 \pi) v^5 + 5\pi v^6 -(5\pi/7)v^7}{160+5 \pi -160 v + 15 \pi v^2 - 5 \pi v^3}~,	~~\text{for}~~~  v<1. 
\end{equation}
Apart from the expressions for average Lanczos coefficients in \eqref{LC_average}, the other
important ingredient used to obtain the above expressions is the two-point energy correlation
function, which, for the GUE, is given by the sine kernel formula. 
We also note that the saturation values for these two cases are respectively given by $1/2$ and $1/3$. These values can be obtained from expressions of $\braket{\mathcal{C}^N_{m}(v)}$ for 
time scales $v \geq 3$ \cite{Kuntal-YiChao}. 

Using these analytical expressions, we now calculate the parameter quantifying the peak that we defined in section \ref{definitionGSC} (see eq. \eqref{peak_parameter}). To determine the location of the peak, we take a derivative of these two expressions with respect to $v$ and find out the lowest value of $v$ (say, $v^m_p$) when the derivative vanishes. Calculating the values 
of the normalized complexity at $v_p$ at this value of the dimensionless time, we determine the 
parameter to be $P_1=0.33$ for spread complexity and $P_2=0.47$ for the second order complexity. An entirely similar procedure can be performed for higher-order complexities 
as well, and we find that $P_3=0.55$, while $P_4=0.60$. We tabulate these values below in the Table. \ref{tab:Pm_contin} for future reference. 

%%%%%%%%%%%%%%%%%%%
%%%%%%%%%%%%%%%%%%%
\begin{table}[h!]
\centering
\begin{tabular}{|c|c|c|c|c|}
\hline
$C_m$ & Peak Value $C_m(t_{\text{peak}})/N^m$& Peak Time $t_{\text{peak}}/N$& Parameter $P_m$ & Ratio $\frac{P_{m+1}}{P_{m}}$ \\ 
\hline
$m = 1$ &0.75& 0.67&0.33 & --- \\ 
$m = 2$ & 0.63&0.71 & 0.47 & 1.42\\ 
$m = 3$ &0.56 &0.74 & 0.55 & 1.17\\ 
$m = 4$ &0.51 &0.76 & 0.60 & 1.09 \\ 
\hline
\end{tabular}
\caption{Numerical values of $\mcc_m(t)$ at the peak, the peak time, and the peak parameter $P_m$  (defined in eq. \eqref{peak_parameter})  for different orders of generalized spread complexity for a GUE Hamiltonian in the continuum limit.}
\label{tab:Pm_contin}
\end{table}
%%%%%%%%%%%%%%%%
%%%%%%%%%%%%%%%%

It is clear from the above discussion that the parameter $P_m$ quantifying the peak in the expressions for the ensemble-averaged complexity has higher values for higher $m$, 
even in the continuum limit, thereby supporting our general claim that the generalized spread complexity of higher orders are more refined probes of the quantum chaotic nature of the spectrum, as far as the spreading 
of an initial time with time in the Krylov basis is concerned. However, it should also be noted 
from the final column of this table, where we have listed the ratio of two successive peak parameters that, after the ratios are gradually decreasing, i.e., looking at a very high order generalized complexity, apart from the first few, one would not necessarily get improved 
characterisation of the peak, which would justify their computation. 

%%%%%%%%%%%%%%%%%%%%%%%%%%%%
\subsection{Generalized spread complexity in sudden quench protocols}
A large part of the present paper deals with the evolution of generalized spread complexity in quantum systems whose thermodynamic equilibrium has been disturbed through an external perturbation. One of the most conventional ways of modeling such non-equilibrium scenarios is through so-called quantum quenches, settings where the parameters appearing in the original system Hamiltonian are assumed to be explicitly dependent on time \cite{Polkovnikov:2010yn}.
In this paper, we consider what is called a sudden
quench protocol, where the system is initially prepared in a state $\ket{\psi_0}$ (which can be an eigenstate of the initial Hamiltonian, $H_0 (\lambda_0)$, or a linear combination of them) and the parameters (collectively denoted as $\lambda_0$)  of $H_0$ are suddenly, and non-adiabatically
changed to a new set of values ($\lambda$) at an instant of time (say, $t_0$). Subsequent
time evolution of the initial state is generated by the new Hamiltonian $H(\lambda)$ for $t>0$. Therefore, one essentially adds an extensive amount of energy to the system through the quench process, and how the system relaxes after such a change can be studied by considering the time evolution of 
different physical quantities, such as different orders of correlation functions, entropy, 
work done on the system or complexity.

In the context of the spreading of an initial state in the Krylov subspace, the sudden 
quench protocols provide a natural set-up to test the robustness of the pattern of the spread 
complexity evolution for many-body chaotic quantum systems that we discussed in the previous subsection. This is related to the fact that in quench set-up, as we have mentioned before, the initial state is constructed from the eigenstate of the pre-quench Hamiltonian, therefore, it is 
completely generic with respect to the post-quench Hamiltonian, which generates the time evolution. To give an example, for the quench protocol involving random matrices that we discuss in the next 
section, the eigenstates of the pre- and post-quench Hamiltonians are completely random 
with respect to each other, as can be verified by computing the so-called inverse participation 
ratio (IPR) 
of the post-quench energy eigenstates in the basis of the pre-quench eigenstates. The IPR of a state $\ket{\psi_0}$ with respect to the basis $\ket{n}$ is defined as\footnote{Here, we adopt the definition of the IPR provided in~\cite{Brandino}, which differs from other definitions commonly found in the literature that correspond to the inverse of the one presented here. For example, see Eq.~(3) in \cite{Buijsman:2023ips}. We thank Pratik Nandy for discussions on this point.} 
\begin{equation}\label{IPR_definition}
    \text{IPR}=\frac{1}{\sum_n |\braket{\psi_0|n}|^4}~.
\end{equation}
In the context of sudden quenches, the IPR of an initial state captures how many eigenstates of post-quench Hamiltonian participate in (i.e., overlap with) the initial state. Based on our definition, the larger the IPR is, the more eigenstates overlap with the initial state. In case the initial state is localized at one post-quench eigenstate, IPR equals one. E.g., for the $\ket{\TFD}$ state defined in eq. \eqref{TFD_post}, the IPR, with respect to the post-quench energy basis, $\ket{n}$ (or with respect to the basis $ \ket{n} \otimes \ket{n} $, if we view TFD state as defined in the doubled Hilbert space) is just equal to $N$.

Here, our goal is to test the generic nature of the peak that appears in the evolution of spread complexity (as well as the generalized spread complexity), which has been used previously as the signature of the chaotic nature of the energy spectrum of quantum many-body systems,  for different generic initial states that appear naturally in the context of a sudden quench protocol. 

The mathematical definitions of the spread and generalized spread complexity remain quite similar to the one in eq. \eqref{GSC_definition}, but the difference from the previous subsection is that,  here 
$H(\lambda)$ denotes the post-quench Hamiltonian, and the Krylov basis elements are constructed by successively applying $H^n(\lambda)$ to the initial state and performing the orthogonalization procedure.  
As we have mentioned above,  the initial state $\ket{\psi_0}$ of the system before the quench is constructed from the eigenstates of $H_0$.    In this paper, we mainly focus on two different such states: an eigenstate of the pre-quench Hamiltonian ($\ket{n_0}$), e.g., the pre-quench ground state and pre-quench infinite temperature TFD state, i.e., TFD constructed from the pre-quench eigenstates, 
\begin{equation}\label{TFD0}
    \ket{\TFD_0}=\frac{1}{\sqrt{N}} \sum_n \ket{n_0} \otimes \ket{n_0}~,
\end{equation}
where $\ket{n_0}$ denotes the pre-quench energy eigenstates. Here, we also mention 
that, unlike the case of $\ket{\TFD}$ state, the auto-correlation function of the state $\ket{\TFD_0}$ for time evolution generated by the post-quench Hamiltonian cannot be interpreted as the spectral form factor. 

To compare the results for the generalized spread complexity and other quantities (such as the IPR),  we also consider the post-quench TFD state as an initial state, even though this case does not correspond to a quench protocol. 

%%%%%%%%%%%%%%%%%%%%%%
%%%%%%%%%%%%%%%%%%%%%%
\section{Sudden quench of random matrices and generalized complexity evolution}\label{RMT_quench}
In this section we consider a sudden quench protocol involving two random matrices from
a one-parameter class of random matrices ($H_r(h)$). First, in section \ref{GOEquench} we take specific values of the parameter $h$ for which both the pre- and post-quench 
Hamiltonians belong to the GOE, while in section \ref{GUEquench} the parameter is chosen in such a way that both Hamiltonians are elements of the GUE. 
Our goal is to study the evolution of the generalized 
spread complexity in such a protocol for different choices of the initial state 
before the quench. Most of the results presented in this section are numerical. An example
of a quench set-up where it is possible to study the evolution of 
generalized spread complexity analytically is discussed in Appendix. \ref{HOquench}. 

\textbf{Quench protocol.}
We consider the following sudden quench protocol, initially proposed for GOE in~\cite{Brandino}. We assume that both the pre and post-quench Hamiltonians are taken from a one-parameter family of $N \times N$ (with $N$ even) random matrices  of the following form 
\begin{equation} \label{denseH}
		H_r(h) = \left(
		\begin{array}{ccc}
			A & h B   \\
			h B^\dagger & C 
		\end{array}
		\right)~.
\end{equation}
Here, the $N/2 \times N/2$ symmetric matrices $A$ and $C$ are sampled  from a normalized random matrix ensemble with measure
\begin{equation}
 \mu(M)  \equiv \exp \Bigg[- \frac{\beta N}{4} \Tr (M^2)\Bigg]~, %~~~\text{with}~~ L=\ln N~,
\end{equation}
where $\beta=1$ for GOE and $\beta=2$ for GUE. The $N/2 \times N/2$ matrix \( B \), with elements \( B_{ij} \), is constructed as follows. For the GOE case, \( B_{ij} \) are real numbers drawn from a normal distribution with zero mean and variance \( 1/N \). For the GUE case, \( B_{ij} \) are complex numbers given by \( B_{ij} = x_{ij} + i y_{ij} \), where both \( x_{ij} \) and \( y_{ij} \) are independently drawn from a normal distribution with zero mean and variance \( 1/(2N) \). 

The Hamiltonians before and after the quench differ from each other in the value of the parameter $h$. In the following, we take $h=-1$ before the quench ($H_0=H_r(-1)$), which is then changed suddenly to a value $h=1$ at $t=0$.  Subsequent time-evolution of the system is generated by this new Hamiltonian $H=H_r(1)$.  
For these choices of the quench parameter $h$, both the pre-quench and post-quench Hamiltonians belong to the GOE or GUE for $\beta=1,2$, respectively, as can be checked by computing the mean and variance of the diagonal and non-diagonal elements of these matrices. 

The Hamiltonian in \eqref{denseH} belongs to a class of Hamiltonians
of the form $H_r(h)=H_0+hV$. Here, the parameter $h$ explicitly breaks a
$Z_2$ symmetry of the initial Hamiltonian $H_0$. This class of Hamiltonians
can also represent some of the 
simplest spin systems, such as the Ising chain in a transverse magnetic field \cite{Brandino}. Since this is a very general class of chaotic Hamiltonians, studying the evolution
of complexity after a sudden quench dynamics can provide us with generic properties of complexity in non-equilibrium chaotic quantum many-body systems. 

We also point out a difference between the random matrices we consider here and the ones constructed in \cite{Brandino}. The authors of \cite{Brandino} assumed the matrices $A$ and $C$ to be GOE random matrices 
taken from a distribution which contains an additional factor of $1/L^2$,
(where $L=\ln N$), in the coefficient of $\Tr (M^2)$, and the matrix $B$  
constructed from a normal distribution with mean zero and variance $L^2/N$. In this paper we have set $L=1$ so that the distribution of eigenvalues obeys the semicircle law $\rho(E)=(2\pi)^{-1} \sqrt{4-E^2}$, for large $N$, while the 
range of the spectrum,  $[-2,2]$, is intensive of the system size. 

\begin{comment}
    The TFD state we are interested in is the infinite temperature version of the usual finite temperature TFD state, 
\begin{equation}
    \ket{TFD}=\frac{1}{\sqrt{N}} \sum_n \ket{E_n} \otimes \ket{E_n} ,
\end{equation}
where $\ket{E_n}$ can be the pre-quench energy eigenstate $\ket{n_0}$ or post-quench eigenstate $\ket{n}$. 
\end{comment}

Next, we present the results of numerical computations for different quantities of interest for this 
quench setup. In our numerical computations, we generate ten independent realizations of the Hamiltonians 
$H_r(-1)$ and $H_r(1)$ with $N=1000$. As mentioned before, we consider two different initial 
states before the 
quench, namely, 
\begin{enumerate}
    \item the pre-quench ground state $\ket{0_0}$,
    \item  the pre-quench TFD state in \eqref{TFD0}. 
\end{enumerate}

Before moving on to consider the time evolution of the generalized spread complexity for different initial states, we first compare the properties of these states 
by computing two quantifiers, both of which are dependent on the post-quench Hamiltonian
generating the evolution, and hence the Krylov basis in a certain way. 
The first of these is the IPR of the initial states with respect to the post-quench energy basis, and the second one is the mean of the distribution of the ratio of two successive Lanczos coefficients.
As we shall see, both of these quantities will provide important indications about the nature of the 
complexity evolution for different initial states that we study in the sequel.

\begin{figure}[h!]
    \centering
    \subfloat[GOE]{
    \centering
        \includegraphics[width=0.48\textwidth]{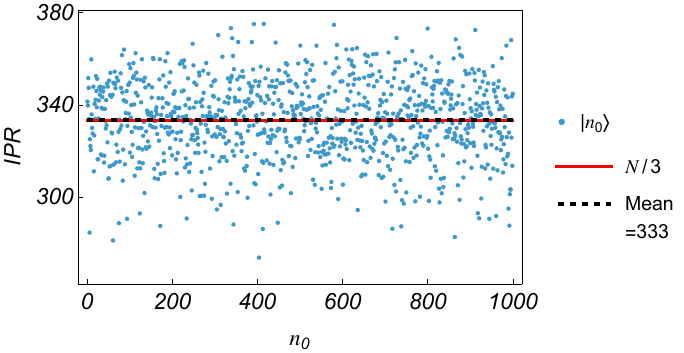} \label{fig:IPRGOE}}
    \hfill
    \subfloat[GUE]{
         \centering
        \includegraphics[width=0.48\textwidth]{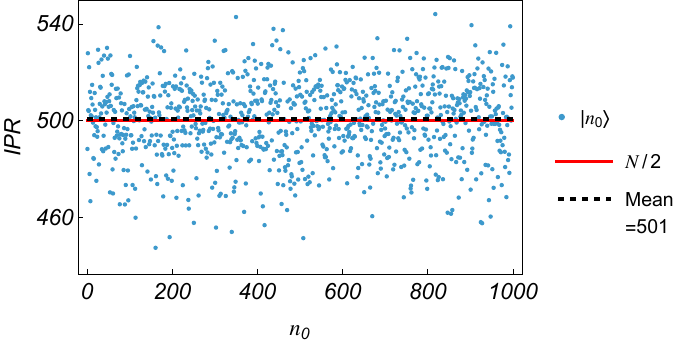} \label{fig:IPRGUE}}
    \caption{Behavior of the IPR for a sudden quench between matrices taken from one realization of GOE (panel (a)) and GUE (panel (b)). The horizontal axis stands for the elements of the pre-quench eigenstates. The blue dots represent the IPRs of pre-quench states $\ket{n_0}$ in the basis of the post-quench eigenstates. The mean values of these distributions are shown with the black dotted lines, which are approximately equal to $N/3$ (shown in red in Fig. \ref{fig:IPRGOE}) for GOE and equal to $N/2$ (red line 
    in Fig. \ref{fig:IPRGUE}) for GUE. The values of the IPRs of the post-quench TFD, pre-quench TFD state, and post-quench ground state in the basis of the pre-quench eigenstates for GOE  are 1000, 331, and 341, respectively. While for GUE, the values IPRs for these three states are, respectively, 1000, 496, and 488.
    %The red solid lines are the values of the IPR of the pre-quench TFD state in the basis of the post-quench eigenstates. The solid brown lines highlight the IPR of the pre-quench ground state. 
    The numerically obtained values of the standard deviations of the distribution of IPRs of pre-quench eigenstates in the basis of the post-quench eigenstates are 17 and 16 for GOE and GUE, respectively.}
    \label{fig:IPR}
\end{figure}

\textbf{IPRs of the pre-quench eigenstates.}
In Fig. \ref{fig:IPRGOE} we have shown the IPR (defined in \eqref{IPR_definition} for a 
generic state $\ket{\psi_0}$) of the pre-quench eigenstates 
$\ket{n_0}$ in the basis of the post-quench eigenstates for a single realization of the GOE matrices. 
As can be seen, the IPRs are almost 
uniformly distributed around the mean value $N/3$, a scaling consistent with one obtained in \cite{Brandino} by using a Porter-Thomas distribution for the eigenvectors. The plot for the IPR clearly indicates that eigenstates are not localized in energy, and pre- and post-quench energy bases are completely random with respect to one another. 
For convenience, in this plot, we have also shown the IPR for the pre-quench TFD state (red line), as well as single out the IPR for the pre-quench ground state (brown line). For this particular
realization, the IPR for 
$\ket{\TFD_0}$ is actually smaller than that of the pre-quench ground state, though it is very close
to the mean value. On the other hand, the IPR for the post-quench TFD state is $N$, much larger 
than that of any typical pre-quench eigenstate. These differences between the IPR values of these states influence the saturation value of the generalized complexity. It is also to be noted that a higher value of IPR for a given initial state indicates that the 
magnitude of the generalized complexity of that state 
would be higher. In that sense, the IPR can be thought of as one of the indicators of whether the initial state
before the quench is a `good' state for observing the 
the peak in complexity, which has been advocated as an indicator of the chaotic nature of the energy spectrum (see the evolution
of complexity that we discuss in section \ref{GOEquench}).

%%%%%%%%%%%%%%%%%
%%%%%%%%%%%%%%%%%
%%%%%%%%%%%%%%%%
%%%%%%%%%%%%%%%

Fig. \ref{fig:IPRGUE} shows the IPRs for the pre-quench eigenstates for a single realization
of the GUE matrix. Here, 
the mean IPR is approximately $N/2$, compared to $N/3$ for the case of the GOE matrix. The IPR
for $\ket{\TFD_0}$ is once again very close to the mean value, though for the realization shown, its value is greater than the IPR for the pre-quench ground state.

\textbf{Comparing the mean of $\tbfr_n= b_{n+1}/b_n$ distribution for different initial states.}
For the one-parameter family of random matrices in \eqref{denseH}, it is interesting to compare the distribution of the ratio of successive Lanczos coefficients for different initial
states. For this class of Hamiltonians with $h=\pm 1$, the Lanczos coefficients are independent 
random variables with known probability distributions. As we discuss in Appendix \ref{Gaussian_beta} below, the distribution of $b_n$s (given in eq. \eqref{bn_distributions}) can be used to obtain the distribution 
of the ratios of successive $b_n$s (which we denote by $\textbf{r}_n= b_{n+1}/b_n$),
as well as that of $\log \tbfr_n$. The resulting analytical form for the distribution of $\tbfr_n$
is given in \eqref{eq-PDF-rn}. 

Our motivation for obtaining the distribution of these quantities comes from the fact that different cumulants of the distribution of $\log \tbfr_n$ have been suggested as the 
indicators of the chaotic or integrable nature of the energy spectrum \cite{Rabinovici:2021qqt, Rabinovici:2022beu, Gautam:2023bcm, Scialchi:2023bmw}. Therefore, as we show below, analytical expressions for the distributions
of these quantities will be helpful in directly comparing these cumulants with the results 
obtained numerically from the Lanczos algorithm. Furthermore, for the quench protocol 
considered here, besides the IPR discussed above, this distribution helps us to quantify the difference between different initial states
prepared before the quench. 

%%%%%%%%%%%%%%%%%%%
%%%%%%%%%%%%%%%%%%
\begin{figure}[h!]
    \centering
    \begin{minipage}{0.45\textwidth}
        \centering
        \includegraphics[width=\textwidth]{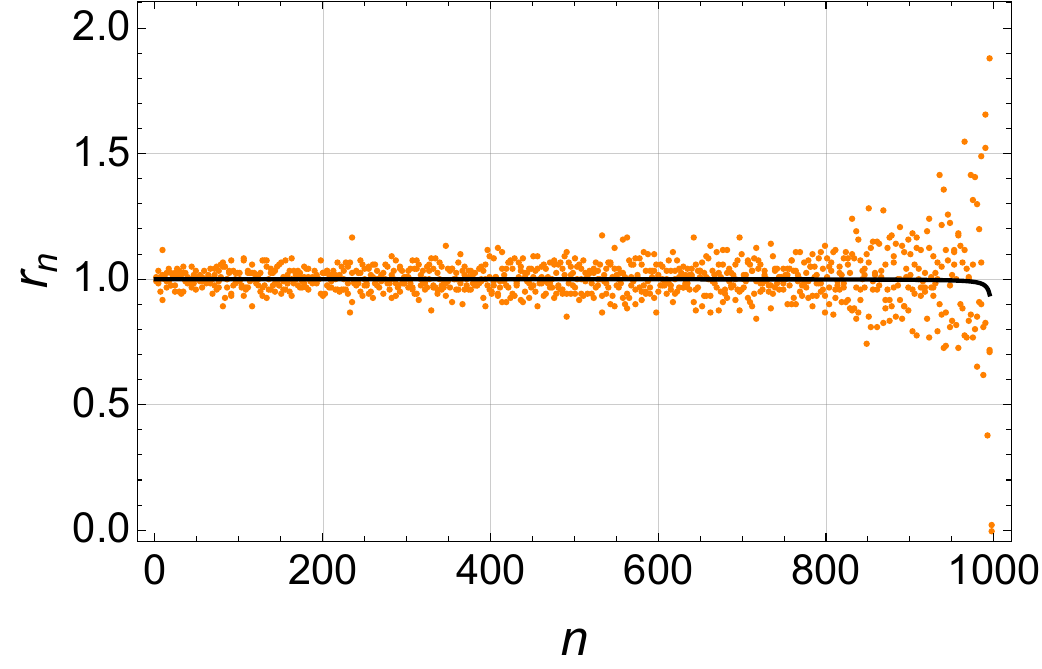}
        \caption*{\small (a) $|\psi_0\rangle=(1, 0, \ldots, 0)^T$} % Caption for the first subplot
        \label{fig:datarn}
    \end{minipage}\hfill
    \begin{minipage}{0.45\textwidth}
        \centering
        \includegraphics[width=\textwidth]{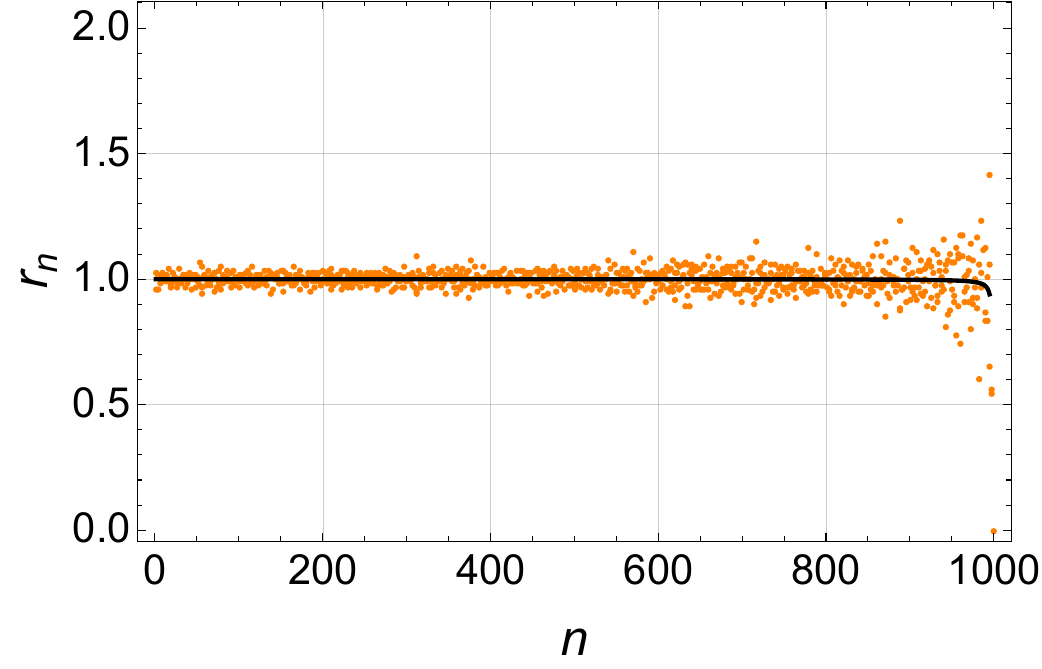}
        \caption*{\small (b) $|\psi_0\rangle=| 0_0\rangle$} % Caption for the second subplot
        \label{fig:datarnA}
    \end{minipage}
    
    \vspace{0.5cm} % Space between rows

    \begin{minipage}{0.45\textwidth}
        \centering
        \includegraphics[width=\textwidth]{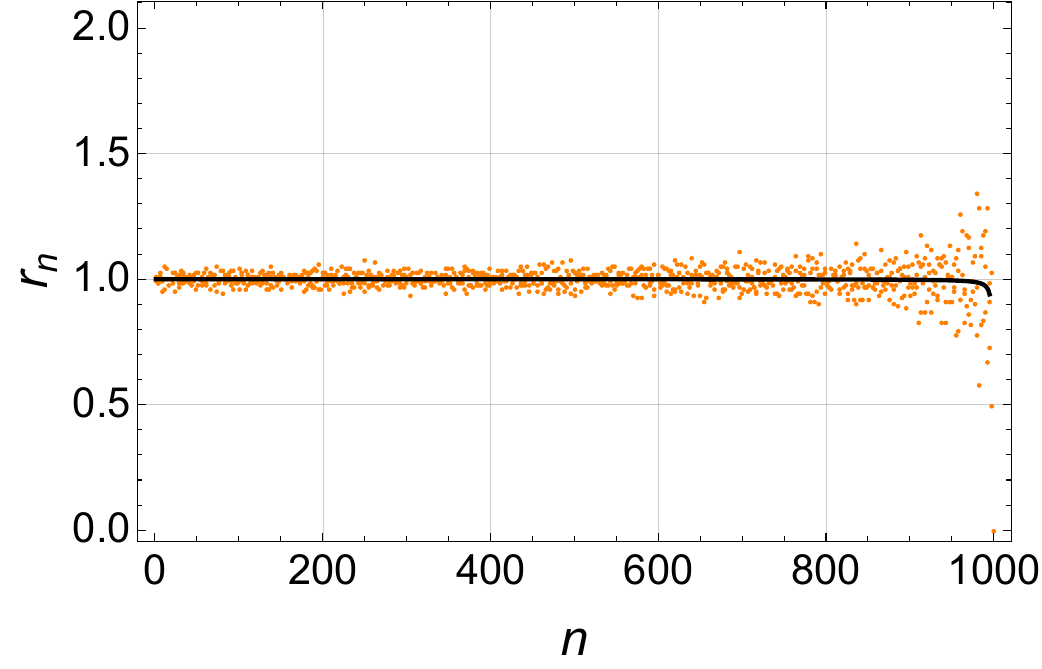}
        \caption*{\small (c) $|\psi_0\rangle=| \TFD_0 \rangle $} % Caption for the third subplot
        \label{fig:datarnB}
    \end{minipage}\hfill
    \begin{minipage}{0.45\textwidth}
        \centering
        \includegraphics[width=\textwidth]{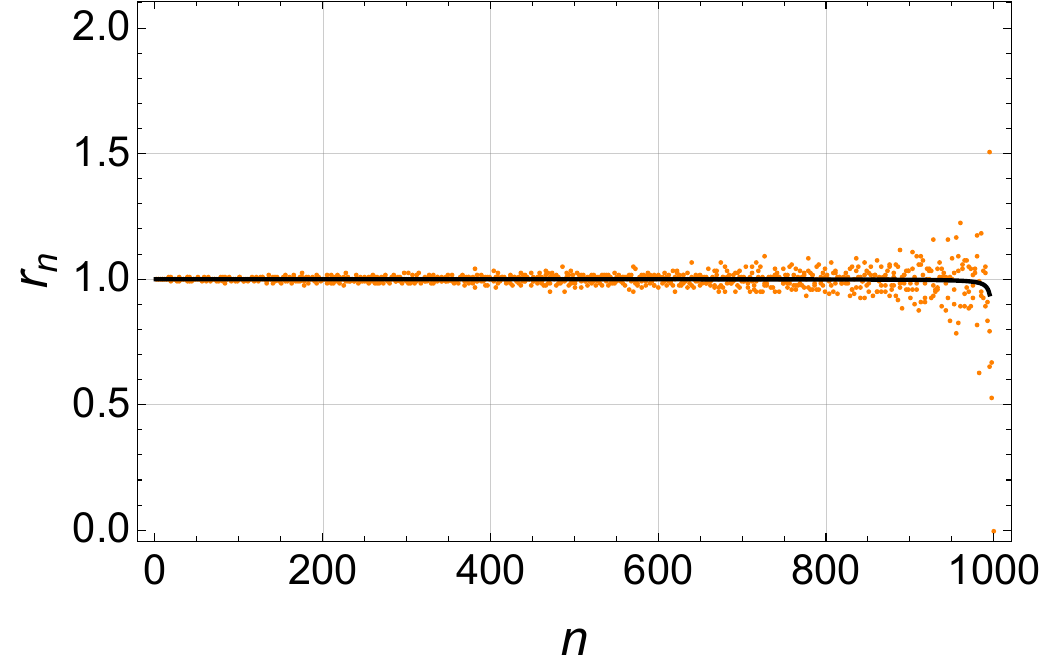}
        \caption*{\small (d) $|\psi_0\rangle=| \TFD \rangle $} % Caption for the fourth subplot
        \label{fig:datarnC}
    \end{minipage}
    \caption{Distribution of the ratio of consecutive Lanczos coefficients, defined as \(\mathbf{r}_n = \frac{b_{n+1}}{b_n}\), plotted as a function of \(n\) for a single realization of a GUE matrix of size \(1000 \times 1000\). The solid black line in all the plots indicates the average value of \(\mathbf{r}_n\), calculated using Equation~(\ref{eq-avgRN}). The initial states $|\psi_0\rangle$ analyzed are: (a) random initial state, \((1, 0, \ldots, 0)^T\); (b) the ground state of the pre-quench Hamiltonian, $| 0_0\rangle$; (c) the TFD state of the pre-quench Hamiltonian, $|\TFD_0 \rangle$; and (d) the TFD state of the post-quench Hamiltonian, $|\TFD \rangle$.}
    \label{fig:four_plots}
\end{figure}
%%%%%%%%%%%%%%%%%
%%%%%%%%%%%%%%%%

In Fig. \ref{fig:four_plots}, we have shown the distribution of $\tbfr_n$ for 
a single realization of a GUE matrix with $N=1000$ for different initial states. The solid black line in each of the plots represents the mean from $\braket{\tbfr_n} $ 
from eq. \eqref{eq-avgRN} with $\beta=2$. Even though the distribution in eq. \eqref{eq-PDF-rn} is valid for a random initial state; it can be seen from Fig. \ref{fig:four_plots} that
the analytical expression for the mean provides quite a good expression for all four 
different initial states. On the other hand, comparing the variance of the distribution of $r_n$s around this 
mean for different initial states, we see that it is quite different in panel (d) to that of (a)-(c). Specifically, for the non-quench scenario,  $\ket{\psi_0}=\ket{\TFD}$, the 
data points for $\tbfr_n$ start to deviate from the mean for relatively larger values of $n$, while for the cases where the initial states correspond to sudden quench, the variation is already quite large for smaller $n$ (specifically in panel (a) corresponding to the pre-quench ground state \footnote{See also the discussion in Appendix \ref{Gaussian_beta} below, where we derive an analytical expression for the variance of 
the $\tbfr_n$ distribution. }. This observation indicates that 
the peak in the generalized spread complexity would be less pronounced for the quenched initial states. 

%%%%%%%%%%%%%%%%%%%
\subsection{Complexity evolution for a sudden quench between two GOE random matrices} \label{GOEquench}

\begin{figure}[h!]
    \centering
    \subfloat[]{
    \centering
        \includegraphics[width=0.48\textwidth]{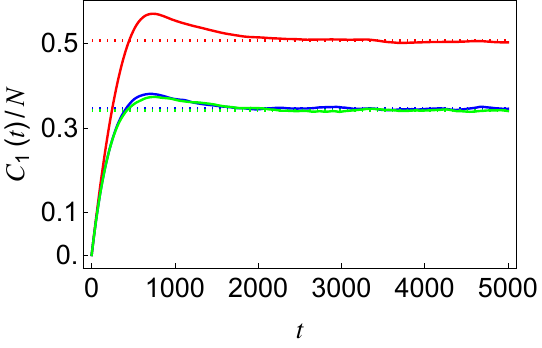} \label{fig: SpreadC-GOE C1}}
    \hfill
    \subfloat[]{
         \centering
        \includegraphics[width=0.48\textwidth]{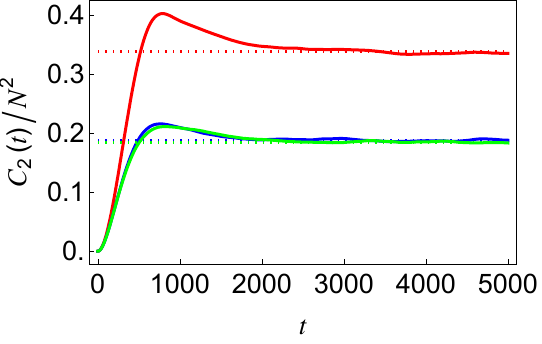} \label{fig: SpreadC-GOE C2}}
        \hfill
        \subfloat[]{
    \centering
        \includegraphics[width=0.48\textwidth]{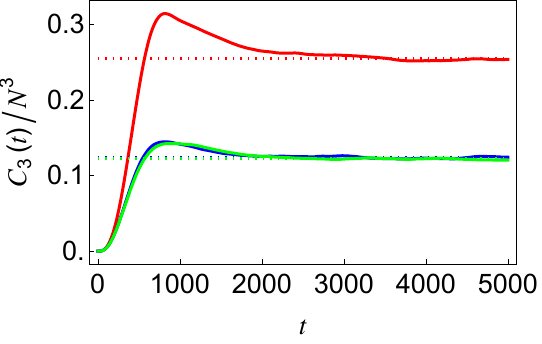} \label{fig: SpreadC-GOE C3}}
    \hfill
    \subfloat[]{
         \centering
        \includegraphics[width=0.48\textwidth]{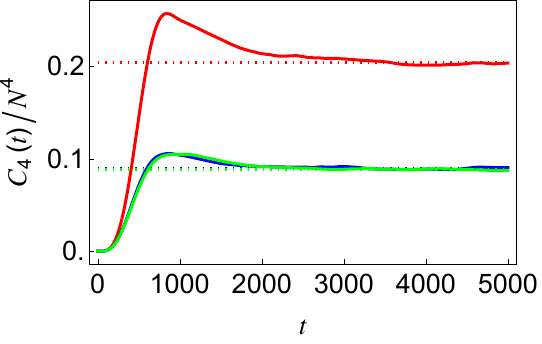} \label{fig: SpreadC-GOE C4}}
    \caption{Time evolution of generalized spread complexities after a sudden  
    quench between two GOE matrices. These are plotted from the average of ten realizations with size $N=1000$. The red, blue, and 
    green curves represent $\mcc_m(t)$, (with $m=1,2,3,4$ in (\ref{fig: SpreadC-GOE C1})-(\ref{fig: SpreadC-GOE C4})),  with respectively $\ket{\TFD}$,
    $\ket{\TFD_0}$ and $\ket{0_0}$ as initial states. The dotted lines in each case indicate the 
    late time saturation values for these quantities are obtained from numerical computations. In each case, we have also normalized $C_m(t)$ by a factor of $N^m$, so that the saturation value of $C_m(t)$ is close to $1/(m+1)$, which is the analytical saturation value for $\ket{\TFD}$ initial state. } 
    \label{fig: SpreadC-GOE}
\end{figure}

The time evolutions of the generalized spread complexity $\mcc_m(t)$ for $m=1,2,3,4$,  
after a sudden  quench between two Hamiltonians
belonging to the GOE are shown in Fig. (\ref{fig: SpreadC-GOE C1})-(\ref{fig: SpreadC-GOE C4}) for two different initial states before the quench (along with a plot for $\ket{\TFD}$ initial state).  In each case, we have also indicated the long-time saturation value of the complexity by dotted lines. The standard deviation around the saturation value is roughly at the third decimal place, which is smaller than $1\%$ of the saturation value. For example, the standard deviation of $C_1(t)/N$ for $\ket{\TFD}$ is 0.0021 with saturation value around 0.35. 

% 0.3453

%%%%%%%%%%%%%%%%%%
%%%%%%%%%%%%%%%%%
\begin{comment}
\begin{figure}[h!]
    \centering
    \subfloat[]{
    \centering
        \includegraphics[width=0.48\textwidth]{GOE_C1.pdf} \label{fig: SpreadC-GOE C1}}
    \hfill
    \subfloat[]{
         \centering
        \includegraphics[width=0.48\textwidth]{GOE_C2.pdf} \label{fig: SpreadC-GOE C2}}
        \hfill
        \subfloat[]{
    \centering
        \includegraphics[width=0.48\textwidth]{GOE_C3.pdf} \label{fig: SpreadC-GOE C3}}
    \hfill
    \subfloat[]{
         \centering
        \includegraphics[width=0.48\textwidth]{GOE_C4.pdf} \label{fig: SpreadC-GOE C4}}
    \caption{Time evolution of generalized spread complexities after a sudden  
    quench between two GOE matrices. They are plotted from the average of four realizations with size $N=1000$. The red, blue and 
    green curves represent $\mcc_m(t)$, (with $m=1,2,3,4$ in (\ref{fig: SpreadC-GOE C1})-(\ref{fig: SpreadC-GOE C4})),  with respectively $\ket{\TFD}$,
    $\ket{\TFD_0}$ and $\ket{0_0}$ as initial states. The dotted lines in each case indicate the 
    late time saturation values for these quantities. In each case, we have also normalized $C_m(t)$ by
    a factor of $N^m$, so that the saturation value of $C_m(t)$ is $1/(m+1)$ for $\ket{\TFD}$ initial 
    state. }
    \label{fig: SpreadC-GOE}
\end{figure}
\end{comment}
%%%%%%%%%%%%%%%%%
%%%%%%%%%%%%%%%%

The overall behavior of $\mcc_m(t)$ with a generic initial state before the quench follows the gradual pattern of rise, peak, slope, and plateau observed for the $\ket{\TFD}$ state in Fig. \ref{fig:c1c2goe1000}. 
Concentrating on the plots for a single order of $\mcc_m$, for different initial states, we see that,
as expected from the observation of the IPR, the complexity is always lower in the case of a quench protocol compared to the non-quench case shown by the red curve, and the magnitude of the peak 
in the former case being quite smaller than the latter case. The early time
behavior of $\mcc_1(t)$ for all three initial states is quite similar; however, for higher order $\mcc_m(t)$s, the curves for different initial states start to deviate before reaching the peak. 
The time scale when $\mcc_m(t)$ reaches a peak seems to be independent of the
initial state. 
Furthermore, for both the pre-quench ground state and pre-quench TFD state, the time evolution of generalized spread complexity is very similar, with their saturation values being approximately equal. 
Now, comparing plots for different orders of $\mcc_m(t)$, one can observe that quadratic growth persists for longer time, and the peak becomes more pronounced for higher orders $\mcc_m$.  
 
Since, one of the main goals of this paper is to understand the effect of the chaotic nature of the spectrum of Hamiltonian in the evolution of generalized spread complexity for a generic initial state, it is important to quantify the peak in more detail, as its presence has been proposed in recent years as a marker of the chaotic or integrable nature of the spectrum. In Tables \ref{tab:GOEPostTFDpar}, \ref{tab:GOEPreTFDpar}, and \ref{tab:GOEpre0par}, we have listed various characteristic quantities 
associated with the plots in Fig. \ref{fig: SpreadC-GOE}, such as the peak value, the saturation 
value and the parameter $P_m$ defined in eq. \eqref{peak_parameter}, as well as the ratio of two successive $P_m$s\footnote{\orange{We have reported all the numerical values in these and the subsequent tables by keeping two significant digits.}}. 

%%%%%%%%%%%%%%%%%%
\begin{table}[h!]
\centering
    \begin{tabular}{|c|c|c|c|c|}
    \hline
        $C_m$ & Peak Value $C_m(t_{\text{peak}})$& Saturation value $\bar{C}_m$& Parameter $P_m$ & Ratio $\frac{P_m}{P_{m-1}}$\\
        \hline
         $m=1$ & 0.57 & 0.50  &  0.12  &   - \\
         \hline
         $m=2$ & 0.40  & 0.34 &  0.15  & \orange{1.3}    \\
         \hline
         $m=3$ & 0.31  & 0.25  &  0.19  & \orange{1.3}   \\
         \hline
         $m=4$  & 0.26  & 0.20 &  0.23   &  \orange{1.2}   \\
         \hline
    \end{tabular}
    \caption{Table of characteristic quantities of generalized spread complexities with $\ket{\TFD}$ as the initial state for the GOE. Parameter $P_m$ is defined in equation (\ref{peak_parameter}).}
    \label{tab:GOEPostTFDpar}
\end{table}

\begin{table}[h!]
\centering
    \begin{tabular}{|c|c|c|c|c|}
    \hline
        $C_m$ & Peak Value $C_m(t_{\text{peak}})$& Saturation value $\bar{C}_m$& Parameter $P_m$ & Ratio $\frac{P_m}{P_{m-1}}$ \\
        \hline
         $m=1$ & 0.38 & 0.35  &  \orange{0.079}  & -  \\
         \hline
         $m=2$ & 0.22  & 0.19 &  0.14  &  \orange{1.8}  \\
         \hline
         $m=3$ &  0.14 &  0.12 &  0.14  &   \orange{1.0} \\
         \hline
         $m=4$  & 0.11  & \orange{0.090} &  0.18  &  \orange{1.3} \\
         \hline
    \end{tabular}
    \caption{Table of characteristic quantities of generalized spread complexities with $\ket{\TFD_0}$ as the initial state for the case of GOE matrices\tablefootnote{We estimated the standard deviation of $P_m$ and found that it is not large enough to make $P_m$ compatible with zero. Additionally, we observe that the magnitude of the peak exceeds that of other background fluctuations, a distinction that can be quantified by computing the parameter $P_m$ for these fluctuations. This suggests the presence of a peak rather than a fluctuation around the saturation value. This observation is particularly relevant when the peak has a small magnitude (mainly in GOE for $P_{m \ge 2}$ with initial states other than post-quench TFD state, see Fig. \ref{fig: SpreadC-GOE}). }. Parameters $P_2$ and $P_3$ only differ at the third decimal place (such that $P_3>P_2$).}
    \label{tab:GOEPreTFDpar}
\end{table}

\begin{table}[h!]
\centering
    \begin{tabular}{|c|c|c|c|c|}
    \hline
        $C_m$ & Peak Value $C_m(t_{\text{peak}})$& Saturation value $\bar{C}_m$& Parameter $P_m$ & Ratio $\frac{P_m}{P_{m-1}}$ \\
        \hline
         $m=1$ & 0.37 & 0.34  & \orange{0.081}   &  -  \\
         \hline
         $m=2$ & 0.21  & 0.19 & \orange{0.095}  &  \orange{1.2}  \\
         \hline
         $m=3$ & 0.14  & 0.12  & 0.14   &  \orange{1.5}\\
         \hline
         $m=4$  & 0.10 & \orange{0.089} &  \orange{0.11}  & \orange{0.79}  \\
         \hline
    \end{tabular}
    \caption{Table of characteristic quantities of generalized spread complexities with $\ket{0_0}$ as the initial state for the GOE.}
    \label{tab:GOEpre0par}
\end{table}
%%%%%%%%%%%%%%%%%%%%%%%

As we can see from the 
fourth column of each of these three tables, the parameter $P_m$ has a value which is higher (or equal to) that of $P_{m-1}$, at least up to $m=3$, for each of the three initial states. This observation strongly supports our general claim that these higher-order generalized complexities can be sharper probes of the chaotic nature of the Hamiltonian
spectrum. 
In this context, a point to note is that for higher order $\mcc_m$s, since the saturation values keep gradually decreasing, the peak parameter is not much larger than the first few $\mcc_m$s.
From the ratio of the peak parameters listed in the final columns of the above tables, 
we also see that $P_4/P_3$ is less than $P_3/P_2$ for two initial states, the $\ket{\TFD}$ and $\ket{0_0}$, while $P_4/P_3$ is larger than  $P_3/P_2$ for $\ket{\TFD_0}$
\footnote{This statement is not true for the quench between GUE Hamiltonians considered below.}. 
%This fact seems to indicate that which of the order $\mcc_m(t)$s is the most efficient indicator of the chaotic nature of the spectrum of GOE Hamiltonian depends on the initial state at the start of the evolution. 
From the numerical values of the ratios
of $P_m$ listed in Tables \ref{tab:GOEPostTFDpar}, \ref{tab:GOEPreTFDpar}, and \ref{tab:GOEpre0par}
we infer that for GOE, the appearance of a peak in $\mcc_2(t)$ and $\mcc_3(t)$ is a more efficient quantifier of the chaotic nature of the energy spectrum than the peak in $\mcc_1(t)$, while the efficiency of the higher order complexities for this purpose depends on the initial state. Also, by this statement, we do not mean to suggest that one should compute only $\mathcal{C}_2$ or $\mathcal{C}_3$, but rather that our numerics indicate these quantities may be more sensitive to the presence of the peak. More generally, our analysis highlights the value of computing $\mathcal{C}_m$ for $m \geq 1$ in any investigation.
 
\begin{figure}[h!]
    \centering
    \subfloat[Dependence of the peak value of $\mcc_1(t)$ and $\mcc_2(t)$ with $N$ for GOE (the brown dashed line is the mean of $C_1^{\text{Peak}}/N$ while the orange dashed line is the mean of $C_2^{\text{Peak}}/N^2$). Each data point for $C_1^{\text{Peak}}/N$ is within the range $0.569 \pm 0.002$. Each data point for $C_2^{\text{Peak}}/N^2$ is within the  range $0.404 \pm 0.004$.]{
    \centering
        \includegraphics[width=0.48\textwidth]{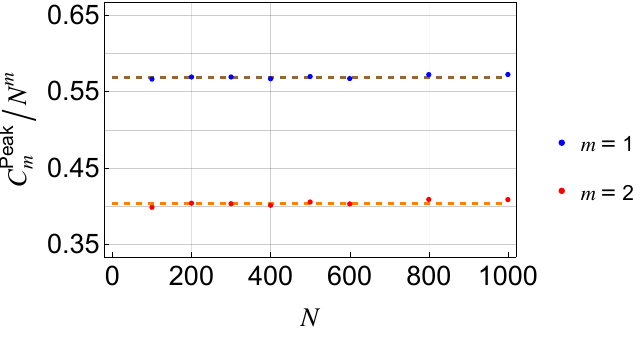} \label{fig:C1C2 Peak GOE-N}}
    \hfill
    \subfloat[Dependence of the peak time of $\mcc_1(t)$ and $\mcc_2(t)$ for GOE (the brown dashed line is the mean of $t_1^{\text{Peak}}/N$ while the orange dashed line is the mean of $t_2^{\text{Peak}}/N$). Each data point for $t_1^{\text{Peak}}/N$ is within the range $0.740 \pm 0.002$. Each data point for $t_2^{\text{Peak}}/N$ is within the range $0.791 \pm 0.007$.]{
         \centering
        \includegraphics[width=0.48\textwidth]{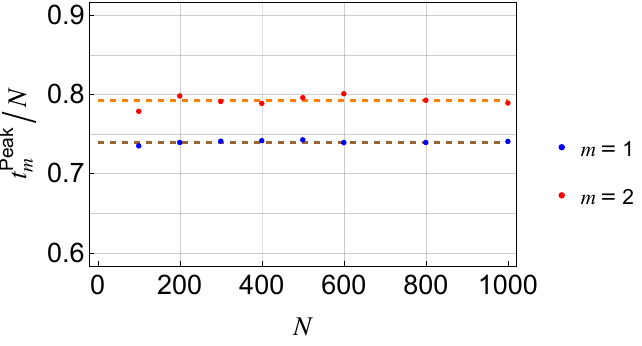} \label{fig:C1C2 Peaktime GOE-N}}
        \hfill
    \subfloat[Dependence of the peak value of $\mcc_1(t)$ and $\mcc_2(t)$ with $N$ for GUE (the brown and orange dashed lines are the same as in (a)). Each data point for $C_1^{\text{Peak}}/N$ is within the range $0.610 \pm 0.002$. Each data point for $C_2^{\text{Peak}}/N^2$ is within the range $0.454 \pm 0.002$.]{
    \centering
        \includegraphics[width=0.48\textwidth]{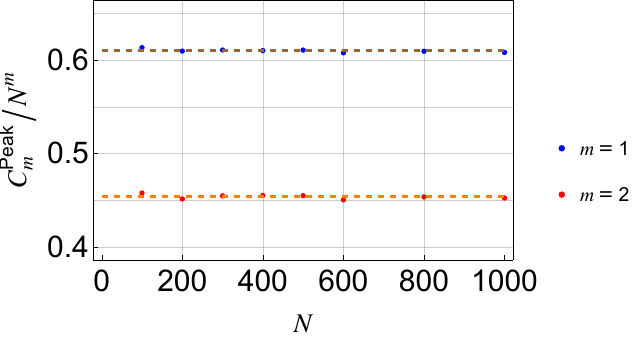} \label{fig:C1C2 Peak GUE-N}}
    \hfill
    \subfloat[Dependence of the peak value of $\mcc_1(t)$ and $\mcc_2(t)$ with $N$ for GUE (the brown and orange dashed lines are the same as in (b)). Each data point for $t_1^{\text{Peak}}/N$ is within range $0.619 \pm 0.004$. Each data point for $t_2^{\text{Peak}}/N$ is within range $0.658 \pm 0.005$.]{
         \centering
        \includegraphics[width=0.48\textwidth]{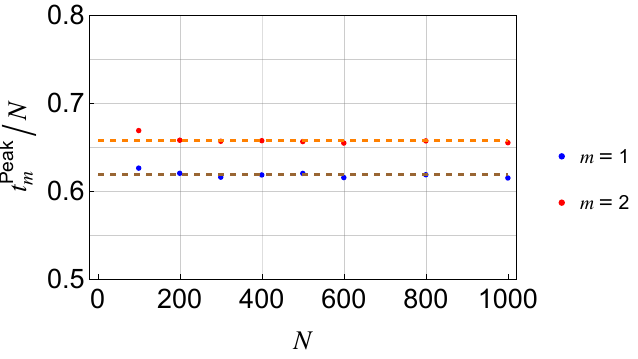} \label{fig:C1C2 Peaktime GUE-N}}
    \caption{Normalized peak value and peak time with N-dependence. Each data point is the average of 20 realizations of GOE/GUE. These results are for the state $\ket{\TFD}$.}
    \label{fig:C1C2GOEGUENdependence}
\end{figure}

\paragraph{$N$-dependence:} To assess the extent to which the above conclusions depend on the size $N$ of the random matrices, we study the peak time $t_\text{peak}$, and the value of the $\mcc_m(t)$ at this time, averaged over 10 realizations of random matrices drawn from a GOE with rank varying from $N=100$ up to $N=1000$ for $m=1,2$. From the results presented in Fig. \ref{fig:C1C2 Peak GOE-N} and \ref{fig:C1C2 Peaktime GOE-N} we observe that $t_\text{peak}/N$ and $C_m(t_\text{peak})/N^m$ are approximately constant as a function of $N$, implying that $t_\text{peak} \sim N$ and $C_m(t_\text{peak}) \sim N^m$. Since the peak parameter, $P_m$, is obtained as a fraction of generalized spread complexities, the dependence on $N$ cancels in this quantity, and hence, it is reasonable to guess that our conclusions would not change if we consider larger values of $N$. These statements remain valid for GUE as can be seen from Fig. \ref{fig:C1C2 Peak GUE-N} and \ref{fig:C1C2 Peaktime GUE-N}. 

%%%%%%%%%%%%%%%%%
%%%%%%%%%%%%%%%%%
\subsection{Complexity evolution for a sudden quench between two GUE random matrices}\label{GUEquench}

\begin{figure}[h!]
    \centering
     \subfloat[]{
    \centering
        \includegraphics[width=0.48\textwidth]{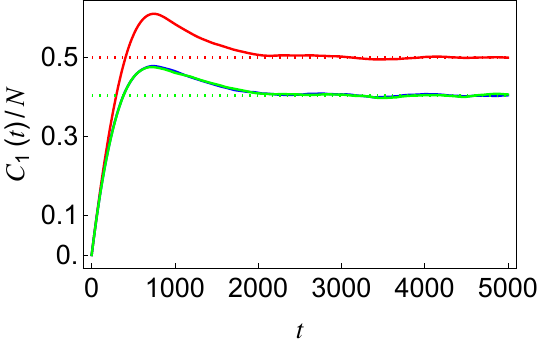} \label{fig: SpreadC-GUE C1}}
    \hfill
    \subfloat[]{
         \centering
        \includegraphics[width=0.48\textwidth]{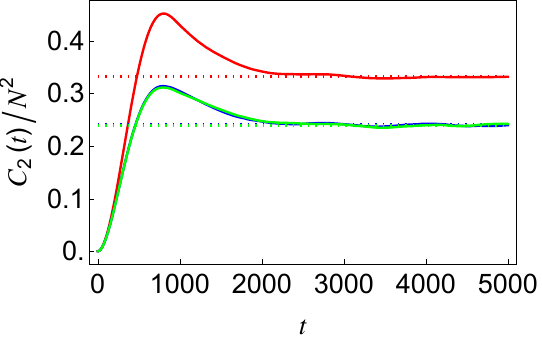} \label{fig: SpreadC-GUE C2}}
        \hfill
        \subfloat[]{
    \centering
        \includegraphics[width=0.48\textwidth]{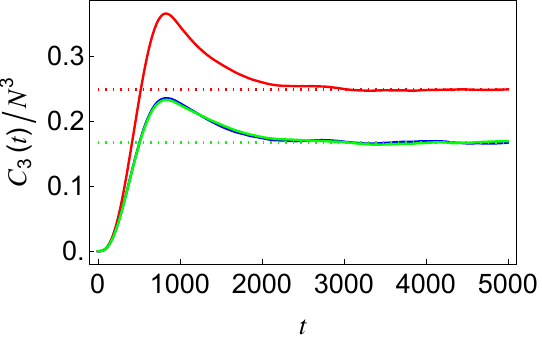} \label{fig: SpreadC-GUE C3}}
    \hfill
    \subfloat[]{
         \centering
        \includegraphics[width=0.48\textwidth]{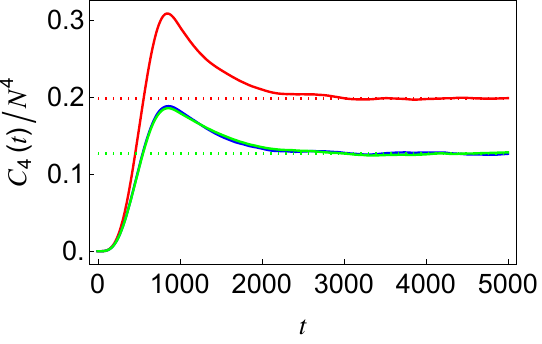} \label{fig: SpreadC-GUE C4}}
    \caption{Time evolution of generalized spread complexities after a sudden  
    quench between two GUE matrices. These are plotted from the average of ten realizations with size $N=1000$. The red, blue and 
    green curves represent $\mcc_m(t)$, (with $m=1,2,3,4$ in (\ref{fig: SpreadC-GUE C1})-(\ref{fig: SpreadC-GUE C4}))  with respectively $\ket{\TFD}$,
    $\ket{\TFD_0}$ and $\ket{0_0}$ as initial states. The dotted lines in each case indicate the numerically  obtained 
    late time saturation values for these quantities. In each case, we have also normalized $C_m(t)$ by
    a factor of $N^m$, so that the saturation value of $C_m(t)$ is close to $1/(m+1)$, which is the analytical saturation value for $\ket{\TFD}$ initial state. }
    \label{fig: SpreadC-GUE}
\end{figure}

We now briefly present the results for quenches between two GUE Hamiltonians. In Figs. (\ref{fig: SpreadC-GUE C1})-(\ref{fig: SpreadC-GUE C4}) we have shown the time evolution of $\mcc_m(t)$, $m=1,..4$, for this case with three different initial states. The general pattern of time evolution once again follows the rise, peak, slope and plateau behavior. However, in this case, the peak 
is more prominent compared to the GOE Hamiltonian for all three initial states and all orders 
of generalized complexity. Once again, the peak and the saturation value of the spread complexity of all orders are almost equal for the pre-quench TFD 
state and the pre-quench ground state. This observation indicates that, as far as the spread complexity (and its higher order generalizations) of time evolution generated by the post-quench Hamiltonian is concerned, these two states are quite similar to each other.   

For convenience, we list the peak parameter for various $C_m(t)$s
for these initial states in Table \ref{tab:GUEpostTFDpar}, \ref{tab:GUEpreTFDpar} and 
\ref{tab:GUE0par}. Clearly, the numerical value of this parameter keeps increasing with order
of $\mcc_m$; however, in contrast to the case GOE matrices, the ratio of two successive parameters keeps decreasing 
for $\ket{\TFD}$ and $\ket{\TFD_0}$ state, while for the pre-quench ground state, it increases after the third order.  This implies that the peaks in $\mcc_2(t)$ and  $\mcc_3(t)$ should efficiently indicate the chaotic nature of the spectrum, irrespective of the initial state.
%implying similar conclusions about the efficiency of $\mcc_2(t)$, and  $\mcc_3(t)$, irrespective of the initial state. 

%%%%%%%%%%%%%%%%%
%%%%%%%%%%%%%%%%
\begin{comment}
\begin{figure}[h!]
    \centering
     \subfloat[]{
    \centering
        \includegraphics[width=0.48\textwidth]{GUE_C1.pdf} \label{fig: SpreadC-GUE C1}}
    \hfill
    \subfloat[]{
         \centering
        \includegraphics[width=0.48\textwidth]{GUE_C2.pdf} \label{fig: SpreadC-GUE C2}}
        \hfill
        \subfloat[]{
    \centering
        \includegraphics[width=0.48\textwidth]{GUE_C3.pdf} \label{fig: SpreadC-GUE C3}}
    \hfill
    \subfloat[]{
         \centering
        \includegraphics[width=0.48\textwidth]{GUE_C4.pdf} \label{fig: SpreadC-GUE C4}}
    \caption{Time evolution of generalized spread complexities after a sudden  
    quench between two GUE matrices. They are plotted from the average of four realizations with size $N=1000$. The red, blue and 
    green curves represent $\mcc_m(t)$, (with $m=1,2,3,4$ in (\ref{fig: SpreadC-GUE C1})-(\ref{fig: SpreadC-GUE C4}))  with respectively $\ket{\TFD}$,
    $\ket{\TFD_0}$ and $\ket{0_0}$ as initial states. The dotted lines in each case indicate the 
    late time saturation values for these quantities. In each case, we have also normalized $C_m(t)$ by
    a factor of $N^m$, so that the saturation value of $C_m(t)$ is $1/(m+1)$ for $\ket{\TFD}$ initial 
    state. }
    \label{fig: SpreadC-GUE}
\end{figure}
\end{comment}
%%%%%%%%%%%%%%%%%
%%%%%%%%%%%%%%%%

%%%%%%%%%%%%%%%%%
%%%%%%%%%%%%%%%%
\begin{table}[h!]
\centering
    \begin{tabular}{|c|c|c|c|c|}
    \hline
        $C_m$ & Peak Value $C_m(t_{\text{peak}})$& Saturation value $\bar{C}_m$& Parameter $P_m$ & Ratio $\frac{P_m}{P_{m-1}}$\\
        \hline
         $m=1$ &  0.61&  0.50 &  0.18  &   - \\
         \hline
         $m=2$ &   0.45 & 0.33&  0.27  &   \orange{1.5}  \\
         \hline
         $m=3$ &  0.37 & 0.25 &   0.32 &  \orange{1.2}   \\
         \hline
         $m=4$  & 0.31  & 0.20 &   0.35  &  \orange{1.1}   \\
         \hline
    \end{tabular}
    \caption{Table of characteristic quantities of generalized spread complexities with $\ket{\TFD}$ as the initial state for the GUE.}
    \label{tab:GUEpostTFDpar}
\end{table}

\begin{table}[h!]
\centering
    \begin{tabular}{|c|c|c|c|c|}
    \hline
        $C_m$ & Peak Value $C_m(t_{\text{peak}})$& Saturation value $\bar{C}_m$& Parameter $P_m$ & Ratio $\frac{P_m}{P_{m-1}}$\\
        \hline
         $m=1$ & 0.48 &  0.40 &  0.17  & -  \\
         \hline
         $m=2$ &  0.31  & 0.24 & 0.23  &  \orange{1.4}   \\
         \hline
         $m=3$ & 0.24 & 0.17 &  0.29 &  \orange{1.3}   \\
         \hline
         $m=4$  & 0.19 & 0.13 & 0.32  &  \orange{1.1}  \\
         \hline
    \end{tabular}
    \caption{Table of characteristic quantities of generalized spread complexities with $\ket{\TFD_0}$ as the initial state for the GUE.}
    \label{tab:GUEpreTFDpar}
\end{table}

\begin{table}[h!]
\centering
    \begin{tabular}{|c|c|c|c|c|}
    \hline
        $C_m$ & Peak Value $C_m(t_{\text{peak}})$& Saturation value $\bar{C}_m$& Parameter $P_m$ & Ratio $\frac{P_m}{P_{m-1}}$\\
        \hline
         $m=1$ & 0.47 &  0.40 &  0.15  &  - \\
         \hline
         $m=2$ &  0.31  & 0.24 & 0.23  &  \orange{1.5}   \\
         \hline
         $m=3$ & 0.23 & 0.17 &  0.26 &  \orange{1.1}   \\
         \hline
         $m=4$  & 0.19 & 0.13 & 0.32  &  \orange{1.2}  \\
         \hline
    \end{tabular}
    \caption{Table of characteristic quantities of generalized spread complexities with $\ket{0_0}$ as the initial state for the GUE.}
    \label{tab:GUE0par}
\end{table}
%%%%%%%%%%%%%%%%%
%%%%%%%%%%%%%%%%

It is also interesting to compare  the values 
of $P_m$ obtained from the numerical computations for GUE presented in Table \ref{tab:GUEpostTFDpar} with the one we obtained analytically from the expressions of $\mcc_m(t)$ in the continuum limit in Subsection \ref{peak_continuum} (listed in Table \ref{tab:Pm_contin}). It can be seen that the continuum limit actually overestimates the values of the peak parameter, even 
though it correctly reproduces the saturation value for the TFD state. It is interesting to 
see whether the approximations in the continuum limit can be improved so that the peak 
parameters obtained from it are more accurate - an issue that we are currently investigating. 

%%%%%%%%%%%%%%
%%%%%%%%%%%%
\section{Summary and discussions}
In this work we have investigated the behavior of generalized spread complexity after sudden quantum quenches within the framework of random matrix theory. Specifically, we considered a sudden quench protocol where the pre-quench Hamiltonian belongs to the GOE or GUE ensembles, and the post-quench Hamiltonian (which also belongs to GOE or GUE) is constructed by dividing the initial Hamiltonian into four equal blocks and flipping the signs of the off-diagonal blocks. For a successful quench, the eigenstates of the pre- and post-quench Hamiltonians must be mutually uncorrelated. We verified this condition by calculating the IPRs of the pre-quench eigenstates in the basis of the post-quench eigenstates, as shown in Fig. ~\ref{fig:IPR}.
   
The quench protocol described above provides a natural framework for studying the state-dependence of generalized spread complexity. Specifically, we numerically analyzed the time evolution of the generalized spread complexities \(C_1\), \(C_2\), \(C_3\), and \(C_4\) for various initial states. These included the well-studied case of the TFD state of the post-quench Hamiltonian, the ground state of the pre-quench Hamiltonian, and the TFD state of the pre-quench Hamiltonian, all evolved under the post-quench Hamiltonian. The results corresponding to an average over ten independent realizations of the random matrices are shown in Figs. \ref{fig: SpreadC-GOE} (for quench between GOE matrices) and \ref{fig: SpreadC-GUE} (for quench between GUE matrices). 

When the initial state is the TFD state of the post-quench Hamiltonian, the generalized spread complexities exhibit a peak just before saturation and then reach a constant equilibrium value from above, mirroring the behavior of spread complexity in chaotic systems as observed in the literature \cite{Balasubramanian:2022dnj, Erdmenger:2023wjg, Camargo:2024deu, Alishahiha:2024vbf}. This behavior contrasts with integrable systems, which are expected to saturate at a constant value from below. The peak is evident for both GOE- and GUE-based quenches. When the initial state is changed to either the ground state or the TFD state of the pre-quench Hamiltonian, a prominent peak is observed in the GUE case, whereas the GOE case shows a milder peak. To quantify the magnitude of these peaks, we define the parameter $P_m=\big(C_m(t_{\text{peak}})-\bar{\mathcal{C}}_{m})/C_m(t_{\text{peak}})$ which measures the relative size of the peak compared to the saturation value of the generalized spread complexities. The results for quenches with different initial states are presented in Tables \ref{tab:GOEPostTFDpar}, \ref{tab:GOEPreTFDpar} and \ref{tab:GOEpre0par} for GOE and Tables \ref{tab:GUEpostTFDpar}, \ref{tab:GUEpreTFDpar}, and \ref{tab:GUE0par} for GUE. In these tables, we report the peak values of the generalized spread complexities $C_m(t_\text{peak})$, their saturation values $\bar{C}_m$, the peak parameter $P_m$, and the ratio $P_m/P_{m-1}$. We observe that the peak parameter increases with increasing $m$, indicating that higher-order generalized spread complexities are more sensitive to the presence of the peak\footnote{It is noteworthy that, for density matrix operators in the SYK model, the generalized Krylov operator complexity grows as $\langle n^m\rangle \sim e^{m \lambda_K t}$ at low temperatures. This observation suggests that higher-order generalized Krylov operator complexities, defined as $\langle n^m\rangle$, may facilitate the numerical detection of the exponential behavior of Krylov operator complexity. We are grateful to Hyun-Sik Jeong for insightful discussions on this point. }. However, the relative increment of $P_m/P_{m-1}$ typically\footnote{For GUE matrices, this is true for the initial states $| \text{TFD} \rangle$ and $| \text{TFD} \rangle_0$, but not for the initial state $| 0_0 \rangle$, where the relative improvement appears to persist at least up to $m=4$, namely $P_4/P_3 > P_3/P_2$.} becomes progressively smaller as $m$ increases, while the saturation value of the generalized spread complexities decreases further. This suggests that the optimal value of $m$ for detecting the presence of the peak for our choices of initial states is likely to be $m=2$ or $m=3$. This numerical analysis is corroborated analytically in Section \ref{peak_continuum} in the continuum limit.
  
We expect our results to capture universal features of sudden quenches in chaotic systems whose spectral statistics align with those of random matrices from GOE or GUE ensembles, depending on the symmetries of the system. It would be interesting to verify this explicitly by studying quantum quenches and generalized spread complexities in realistic chaotic quantum many-body systems. Additionally, it would be valuable to investigate the corresponding features in systems that are neither strongly chaotic nor integrable, whose spectral properties can be approximately modeled by the so-called Gaussian $\beta$-ensembles. We expect to report on this in the near future. 

Also note that, based on our numerical results in Fig. \ref{fig:C1C2GOEGUENdependence}, the peak value and peak time for any order $m$ are linear in the rank of the Hamiltonian $N$ with different pre-factors. This observation remains valid for GOE and GUE. We believe this statement should hold for any $\beta$-ensemble, which can also be generalized to real physical systems belonging to a Gaussian $\beta$-ensemble with specific values of $\beta$. It would be very interesting to test the validity of this linear dependence on $N$ for other physical systems, such as the SYK model. Moreover, it would be interesting to prove this statement using analytical computations.
    
Finally, we comment on recent developments regarding the study of Krylov complexity in holography. The rate of Krylov complexity of quantum states excited by local operators in conformal field theories was recently found to be connected with the radial momentum of particles in Anti-de Sitter spacetimes~\cite{Caputa:2024sux, Fan:2024iop, He:2024pox}. It would be interesting to understand if a similar statement could be made for generalized spread complexities.

%%%%%%%%%%%%%
%%%%%%%%%%%%%%

\acknowledgments
We would like to thank Alexander Altland, Matteo Baggioli, Xiangyu Cao, Pawel Caputa, Antonio Garcia-Garcia, Alexander Jahn, Hyun-Sik Jeong, Rene Meyer, Pratik Nandy, Kunal Pal, Maxim Pavlov, Giuseppe Policastro, and Shozab Qasim for valuable discussions.  We sincerely thank the referee for various useful comments on a draft version of the manuscript.
This work was supported by the Basic Science Research Program through the National Research Foundation of Korea (NRF) funded by the Ministry of Science, ICT $\&$ Future Planning (NRF-2021R1A2C1006791),  by the Ministry of Education (NRF-2020R1I1A2054376) and the AI-based GIST Research Scientist Project grant funded by the GIST in 2025.
This work was also supported by Creation of the Quantum Information Science R$\&$D Ecosystem (Grant No. 2022M3H3A106307411) through the National Research Foundation of Korea (NRF) funded by the Korean government (Ministry of Science and ICT). H.~A. Camargo and V.~Jahnke were supported by the Basic Science Research Program through the National Research Foundation of Korea (NRF) funded by the Ministry of Education (NRF-2022R1I1A1A01070589 and RS-2023-00248186). H. Camargo, Y. Fu, V. Jahnke, and K. Pal should be recognized as co-first authors.

%%%%%%%%%%%%%%%%%%%%%%%
%%%%%%%%%%%%%%%%%%%%%%%%%

\appendix
%%%%%%%%%%%%%%%%%%%%%
%%%%%%%%%%%%%%%%%%%%
\section{Quench between two harmonic oscillators: Generalized spread complexity evolution}\label{HOquench}
In this Appendix, we discuss a specific sudden quench set-up for which it is possible to obtain 
the expressions for the generalized spread complexity and other relevant quantities analytically. The protocol we consider is the following: we assume a modified version of the usual form  of the spreading operator to be  the Hamiltonian before the quench, i.e.,
\begin{equation}
     H_0(\omega_0)= \omega_0 \sum_n \Big(n+\frac{1}{2}\Big)  \ket{K_n}\bra{ K_n} = \omega_0 \sum_n \Big(n+\frac{1}{2}\Big)  \ket{n_0}\bra{ n_0} ~.
\end{equation}	
Here, $\ket{n_0}$ denote the eigenstates of the pre-quench Hamiltonian, which according to our protocol are the Krylov basis associated with the post-quench Hamiltonian  $H(\omega)$, i.e., 
$\ket{n_0}=\ket{K_n}$. Here
 $H(\omega)$ is assumed to be of the form, $H(\omega)= \omega \sum_n (n+\frac{1}{2})  \ket{n}\bra{ n} $. If we set $\omega_0=1$, and neglect the ground state energy, 
the pre-quench Hamiltonian reduces to the standard spreading operator in the Krylov basis.
Thus, the protocol essentially describes a quench 
between two harmonic oscillators with different frequencies (we have assumed the mass of both oscillators to be unity for convenience).  The initial state
before the quench is the ground state of the initial Hamiltonian, i.e., $\ket{\psi_0}=\ket{0_0}=\ket{K_0}$.  For such a protocol, it is possible to compute all the relevant quantities analytically. 
	
To compute the transition probabilities in the Krylov basis, $p_n(t)= |\phi_n(t)|^2 $, we need the  expressions for the overlap between the pre- and post-quench eigenstates, i.e., 
those between the Krylov basis and the new energy eigenstates. 
These expressions for two harmonic oscillator Hamiltonians are quite well known in the literature, see, e.g.,  \cite{Sotiriadis} and \cite{Palma}.  
The general formula for the overlap between eigenstates of different oscillators is 
	\begin{equation}\label{overlap}
		\begin{split}
			\braket{m_0|n}= (m!n!)^{1/2}\braket{0_0|0} \Bigg(\frac{\Lambda}{2}\Bigg)^{(m+n)/2} ~
			\times \sum_{k=0}^{[m,n]} \frac{(-i)^{m-k}}{k! (m-k)!(n-k)!} ~\Bigg(\frac{\nu}{2}\Bigg)^{-k} ~H_{m-k}(0) H_{n-k}(0)~,
		\end{split}
	\end{equation}
where we have denoted, 
	\begin{equation}
		\mu=\frac{1}{2} \frac{\omega_0+\omega}{\sqrt{\omega \omega_0}}~,~\nu=\frac{1}{2} \frac{\omega_0-\omega}{\sqrt{\omega \omega_0}}~,~~\text{and}~~
		\Lambda=\nu/\mu=\frac{\omega_0-\omega}{\omega_0+\omega}~.
	\end{equation}
Also, $\braket{0_0|0} $ denotes the overlap integral between the two ground states, which one can evaluate straightforwardly by using the formula for the 
ground state wavefunction of the two oscillators to be, $\braket{0_0|0}=1/\sqrt{\mu} $, and $H_n(x)$ denote  the Hermite polynomials. 
We also note that the sum in the above expression \eqref{overlap} is finite and the upper limit $[m,n]$ denotes the smaller of $m$ or $n$. 
As a special case of this general formula, one can derive the expression for the overlap between the initial state and the post-quench energy states to be 
	\begin{equation}\label{overlap2}
		\braket{0_0|2n} = (-1)^n \frac{\sqrt{(2n)!}}{\sqrt{\mu}n!} \Bigg(\frac{\Lambda}{2}\Bigg)^n~.
	\end{equation}
This overlap is non-zero only for even states, i.e., $	\braket{0_0|2n+1}=0 $.  
	
\begin{figure}[h!]
		\centering
		\includegraphics[width=3in,height=2.2in]{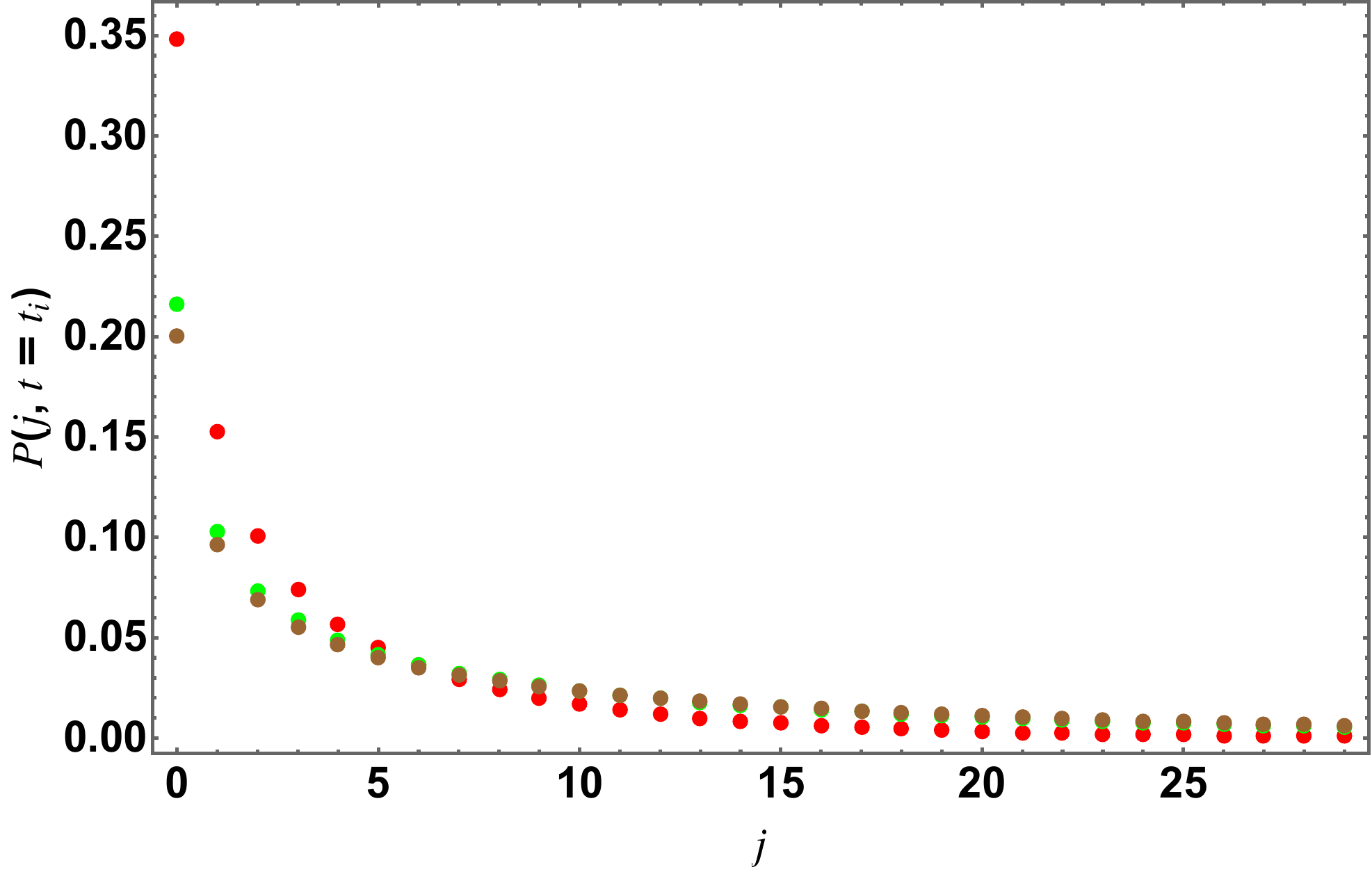}
		\caption{The distribution of the spreading number $P(j,t=t_i)$ for a quench between two harmonic oscillators, where the pre-quench oscillator is the spreading operator in the post-quench Krylov basis. The fixed values of time are $t=1$ (red), $t=2$ (green), $t=3$ (brown), and the frequencies before and after the quench are, respectively, $\omega_0=1$, $\omega=10$. }
		\label{fig:sodistribution}
	\end{figure}

First, we compute the transition probabilities, $p_n(t)= |\phi_n(t)|^2 $, which in terms of the post-quench energy basis can be rewritten as (with $\rho_0=\ket{0_0}\bra{0_0}$)
	\begin{equation}\label{pnho}
		p_n(t)= \sum_{k,j}  \braket{j|n_0} \braket{n_0|k} \braket{k|\rho_0|j}e^{-i(k-j)\omega t}~. 
	\end{equation}
In both the sums, only non-zero contributions come from even values of the indices. 
Using the expressions for the overlaps in \eqref{overlap2} to compute the matrix elements of the initial state  density matrix, we get
	\begin{equation}
		p_n(t)= \frac{1}{\mu}\sum_{k,j} (-1)^{(j+k)} \frac{\sqrt{(2j)!(2k)!}}{j! k!}~ \Bigg(\frac{\Lambda}{2}\Bigg)^{j+k}~\braket{2j|n_0} \braket{n_0|2k} ~e^{-2i(k-j)\omega t}.
	\end{equation}
Here, the sum is over all the integers.   Instead of performing  the above summation explicitly to compute $p_n(t)$, and subsequently, the 
expression for the generalized spread complexity,  it is more convenient to use the fact that, in terms of the pre-quench creation and annihilation operators, the form of the post-quench Hamiltonian is such that it is essentially an element of the $su(1,1)$ Lie algebra. 
Using the standard decomposition formulas for the $SU(1,1)$ group elements, it is then straightforward to derive the expression for the complexity and other higher-order
moments of the spreading.  The analytical formula for the generating function \eqref{generating} is given by
	\begin{equation}
		G(\eta,t)= \bigg[\cos^2 (\omega t)+ \Big(\Omega^2 - e^{\eta} \tilde{\Omega}^2\Big) \sin^2(\omega t)\bigg]^{-1/2}~,
	\end{equation}
where we have denoted
	\begin{equation}
		\Omega= \frac{\omega_0^2 +\omega^2}{2 \omega \omega_0}~,~~\tilde{\Omega}= \frac{\omega_0^2 - \omega^2}{2 \omega \omega_0}= \mu \nu ~.
	\end{equation}
Taking a derivative of $G(\eta,t)$ with respect to $\eta$ and subsequently setting $\eta=0$ we get the expression for the mean of the spreading, i.e., the spread complexity\footnote{Note that this expression for the 
spread complexity is not exactly equal to the expectation value of the modified spreading operator, $H_0 (\omega_0$) with respect to the time-evolved state due to the zero point energy. The generating function, as well as the higher order moments, are calculated with respect to the usual spreading operator $\mck$. } \cite{spread2},
	$\mcc= \frac{1}{2}\tilde{\Omega}^2 \sin^2(\omega t)$.  Similarly, from the higher-order moments of the generating function, one can obtain the expressions for the 
	higher-order generalized complexity. E.g., the expression for the variance of the  spreading is given by
	\begin{equation}
		\Big(\Delta n(t)\Big)^2 = \frac{1}{32} \tilde{\Omega}^2 \Big(\Omega^2 \sin^4 (\omega t) + \sin^2 (2 \omega t)\Big)~. 
	\end{equation}
Analytically continuing the generating function to the complex value of the parameter, and subsequently taking a Fourier transform, we obtain the probability distribution of the spreading number \cite{Kuntal-YiChao}. 
 This distribution of the spreading number for the  quench protocol under consideration, for a fixed value of time, $t=t_i$,  therefore contains a sequence of delta function  peaks of decreasing heights corresponding to decreasing transition probabilities between the
 Krylov basis and the eigenbasis of the post-quench Hamiltonians. These peaks are equidistant since the spreading operator eigenvalues are also equidistant. 
As the value of the time is increased, the height of these peaks starts to decrease for the first few values of $j$. However, as $j$ is increased further, there is a crossover between the strength of the peaks due to revivals in the transition probabilities.  A plot of the distribution $P(j,t=t_i)$ for three different fixed values of time  $t_i$ is shown in Fig. \ref{fig:sodistribution}. 

%%%%%%%%%%%%%%
%%%%%%%%%%%%%%
%%%%%%%%%%%%%%
%%%%%%%%%%%%%%

%%%%%%%%%%%%%%%%%%%%%%%%
%%%%%%%%%%%%%%%%%%%%%%%%%
\section{Gaussian $\beta$-ensemble and distribution of ratios of successive $b_n$s}\label{Gaussian_beta}
\textbf{Gaussian $\beta$-ensemble.}
In this appendix we derive the probability distribution for the ratio of two successive Lanczos coefficients of the second type ($b_n$) for a random matrix Hamiltonian $H_{\beta}$ drawn from the Gaussian $\beta$-ensemble,
also commonly known as the $\beta$-Hermite ensemble \cite{dumitriu2002matrix, forrester2010log}. Here 
 $\beta$ is a continuous parameter that can take values between $0$ and infinity and is known as the Dyson index of the ensemble. The cases $\beta=1,2,4$ are well-known GOE, GUE and GSE, respectively, while taking the limit $\ \beta\rightarrow 0$, one can approximately match with the Poisson level statistics typically observed in integrable systems as well. An example of a physical system whose energy level spacing statistics shows a good level of agreement with that of the Gaussian  $\beta$-ensemble is a spin model 
that has a many-body localized phase. In \cite{buijsman2019random} such a spin model, namely a spin 1/2-XYZ chain with disorder was studied, and it was shown that both the statistics of the nearest eigenvalue spacing and that of the so-called $r$-parameter of this particular model approximately matches with the Gaussian $\beta$-ensemble for all ranges of the disorder parameter.\footnote{We however also note that, the proposal that the $\beta$-ensembles can model spin 1/2 Heisenberg chains was shown not to be supported by analysis of other relevant quantities, such as the statistics of higher-order 
spacing of the energy levels, the ratio of nearest-neighbour spacings, and the spectral form factor. See refs. \cite{sierant2019level, sierant2020model, relanobeta} }

Here we consider a random matrix Hamiltonian $H_{\beta}$ belonging to the Gaussian generalized $\beta$-ensemble. Thus $H_{\beta}$ is  a
random real symmetric tridiagonal $N \times N$ matrix for which the 
distribution of the diagonal and non-diagonal elements can be  schematically denoted  as,  
$\frac{\mathcal{N}(0,2)}{\sqrt{\beta N}}$ and $\frac{\chi_{(N-n)\beta}}{\sqrt{\beta N}}$, respectively.
Hence, essentially, the diagonal elements are independent Gaussian random variables with 
mean $0$ and variance $2$, while the upper(or lower)-diagonal elements are drawn from independent chi distributions. 
Since the matrix is in the tridiagonal form, these distributions also fix the distribution
of the Lanczos coefficients associated with $H_{\beta}$. Therefore, %\red{for a random initial state}, 
the first set of Lanczos coefficients $a_n$s are independent Gaussian random variables
with zero mean and variance $2/N$\footnote{For a matrix drawn from GOE, this variance is the same as the variance of the diagonal Gaussian random variables.}, while the second set of 
Lanczos coefficients ($b_n$) are independently distributed with the following distribution 
\cite{Trotter1984EigenvalueDO,dumitriu2002matrix}
\begin{equation}\label{bn_distributions}
    p(b_n)= \frac{2}{\Gamma (k/2)} \Bigg(\frac{\beta N}{2}\Bigg)^{k/2} b_n^{k-1} e^{-\beta N b_n^2/2}~,
\end{equation}
where $k=(N-n)\beta$. For our later purposes, it is useful to record the expressions for the averages of these distributions, specifically in the large $N$ limit. In this limit (and 
for $\beta>1$) the averages can be approximated  by\footnote{The average of the distribution of $a_n$ is not an approximation; it is always zero, irrespective of the values of $N$ and $n$.} \cite{Balasubramanian:2022dnj,Erdmenger:2023wjg}
\begin{equation}\label{LC_average}
    \braket{a_n}=0~,~~~\text{and}~~~ \braket{b_n}= \sqrt{1-\frac{n}{N}}~. 
\end{equation}
We also note that variances of both the distributions of $a_n$s and $b_n$s scale as $1/N$ in the leading order.

%\red{random initial state for GOE and GUE ..only?}

The Hamiltonian $H_\beta$ can be thought of as representing 
a one-dimensional lattice having open boundary conditions and is highly inhomogeneous. 
A particle on this lattice can randomly hop to its nearest neighbors under disordered 
on-site potentials. The inhomogeneity comes from the fact that, on average, non-
diagonal elements of $H_\beta$ change with $n$ as one moves along the lattice. 
It has been observed recently that the $\beta$-ensemble has three different phases as one changes $\beta$  \cite{Das:2021hmt}. 
Reparametrizing the Dyson index as $\beta=N^{-\gamma}$, these phases are, respectively, an ergodic phase (for $\gamma \leq 0$), a nonergodic extended phase (when $0<\gamma<1$), and a localized phase (for $\gamma \geq 1$). Hence, an ergodic transition (as well as a chaotic-integrable transition) occurs at $\gamma=0$, while an Anderson transition occurs at $\gamma=1$.

\textbf{Distribution of the ratios of successive $b_n$s.}
Consider the distribution of the quantities $\textbf{r}_n=\frac{b_{n+1}}{b_n}$, where $n=0, \cdots N-1$.   Using the formula for the ratio distribution, we obtain the probability
distribution function for the $\textbf{r}_n$ to be of the following form (with $k=(N-n) \beta$) 
\begin{equation} \label{eq-PDF-rn}
    p(\textbf{r}_n)= N_n ~\tbfr_n^{k-\beta-1} \big(\tbfr_n^2+1\big)^{\frac{\beta}{2}-k}~,~~
    \text{with}~~ N_n= \frac{2 \Gamma\Big(k-\beta/2\Big)}{\Gamma \Big((k-\beta)/2\Big) ~\Gamma\Big(k/2\Big)}~. 
\end{equation}
It is easy to calculate the first few moments of this distribution analytically. E.g., the 
expressions for the mean and the second-order moment of the distribution are given, respectively,  by \footnote{We note that $\braket{\tbfr_n}$ represents the mean value of the distribution of $\tbfr_n$ for a fixed value of the index $n$ on the Krylov chain, \textit{not} the average over different values of $n$. In this regard, this should be distinguished from the standard $r$-parameter studied in the context of the random matrices, which we discuss in the next section (see eq. \eqref{r_param_ave}.  In particular,  the numerical value of $\langle \tilde{r} \rangle$ measures the amount of correlation between nearest-neighbor (unfolded) energy levels, whereas $b_n$ Lanczos coefficients are independently distributed random variables by construction of the tridiagonal matrix form.}
\begin{equation} \label{eq-avgRN}
    \braket{\tbfr_n} = \frac{\Gamma\Big((k-1)/2\Big) \Gamma\Big(\big(k-\beta+1\big)/2\Big)}{\Gamma\Big((k-\beta)/2\Big) \Gamma\Big(k/2\Big)}~, ~~\text{and}~~
    \braket{\tbfr_n^2} = \frac{\Gamma\Big((k-\beta)/2 + 1\Big) \Gamma\Big(k/2-1\Big)}{\Gamma\Big((k-\beta)/2\Big) \Gamma\Big(k/2\Big)}~. 
\end{equation}
When $n=N-1$, the expression for the mean reduces identically to zero, which is consistent with the fact that the last
$b_n$ at the end of the Krylov chain vanishes. 
From these two expressions one can calculate the variance ($(\Delta \tbfr_n)^2 = \braket{\tbfr_n^2} - \braket{\tbfr_n}^2 $) of the distribution as well.  
Note that, in general, similar to the distribution of $b_n$, both the mean and variance are functions $n, N$ and the Dyson index $\beta$. 
However, from the expression for $\braket{\tbfr_n}$ it can also be checked that for a Hamiltonian sampled from the GOE ($\beta=1$), the mean actually becomes independent of both $N$ and $n$ and has a constant value 1. 
%Furthermore, it can also be checked that in the limit $N \rightarrow$
%the mean becomes independent of $\beta$ and $n$ (as in the distribution of $b_n$) and takes a constant value 1. 
Furthermore, performing an expansion of $\braket{\tbfr_n} $ and $\braket{\tbfr_n^2} $ as a function of $1/N$ with fixed $\beta$, we obtain
\begin{equation}
    \braket{\tbfr_n} \approx 1-\frac{1}{2N} \Big(\frac{1}{\beta} -1\Big) + \mathcal{O}(1/N^2)~, ~~\text{and}~~
    \braket{\tbfr_n^2} \approx 1-\frac{1}{N} \Big(\frac{2}{\beta} -1\Big) + \mathcal{O}(1/N^2)~. 
\end{equation}
Therefore, in the limit of large $N$, at the leading order ($\mathcal{O}(1))$ the mean of the distribution of $\tbfr_n$
is close to $1$, irrespective of $\beta$ and $n$, while the variance, in the leading order 
can be approximated by $\Delta^2 (\tbfr_n) \approx \frac{1}{\beta N}+ \mathcal{O}(1/N^2)$. 
On the other hand, taking the limit $\beta \rightarrow 0$ with fixed finite $N$ (i.e., $\gamma \gg 1$ and the system is deep in the localized phase),  it can be seen that in the limit when the level spacing distribution of the  Gaussian $\beta$-ensemble gets close to the Poisson distribution,  the variance of the distribution of successive $b_n$s indeed diverges. 
We also note that when both the limits $\beta \rightarrow 0$, and $N \rightarrow \infty$ are taken simultaneously (i.e., $\gamma>0$, and the system is in either the nonergodic or the localized phase), while keeping their product $x= \beta N$ fixed, the variance becomes 
independent of $n$ and has an expansion of the form $\sum_j (c_j/x^j)$, where $c_j$ are
constants independent of the Krylov basis index $n$. This limit is also sometimes referred to as the
quantum mechanics limit of the $\beta$-ensemble in the literature \cite{Krefl:2013bsa}. 

\begin{figure}[h!]
    \centering
    \begin{minipage}[b]{0.32\textwidth}
        \centering
        \includegraphics[width=\textwidth]{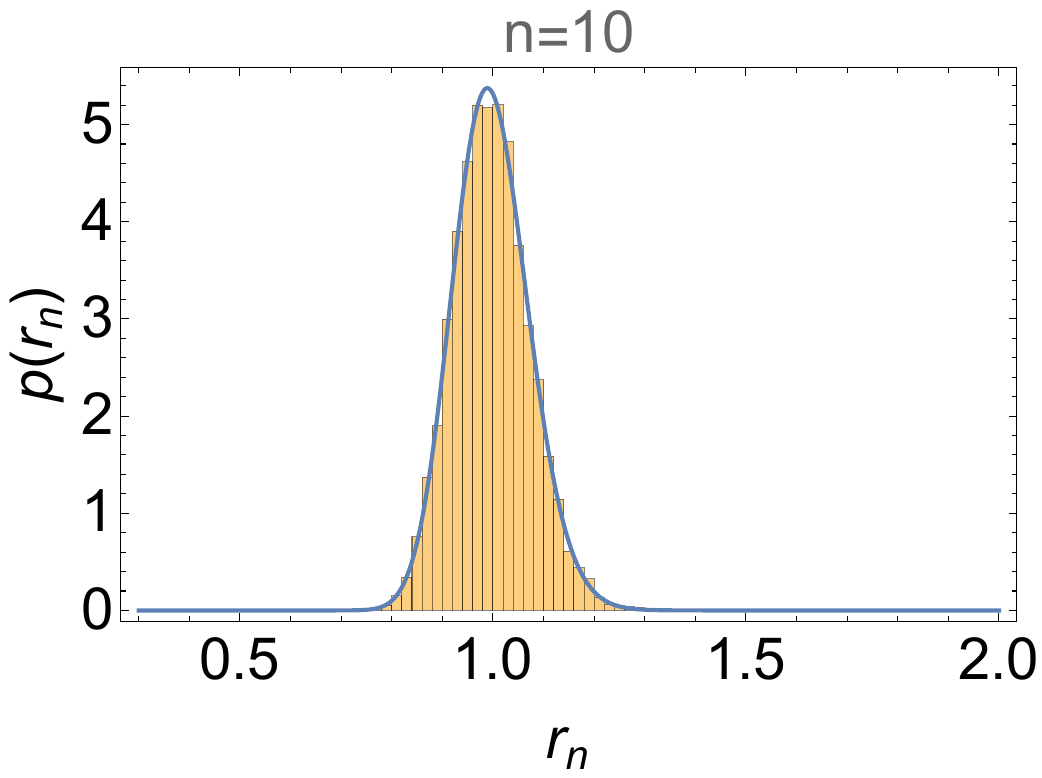} % Replace with your image file name
        %\caption*{Left: $n=10$} % Optional: subfigure caption without numbering
        %\label{fig:fig_a} % Label for left panel
    \end{minipage}
    \hfill
    \begin{minipage}[b]{0.33\textwidth}
        \centering
        \includegraphics[width=\textwidth]{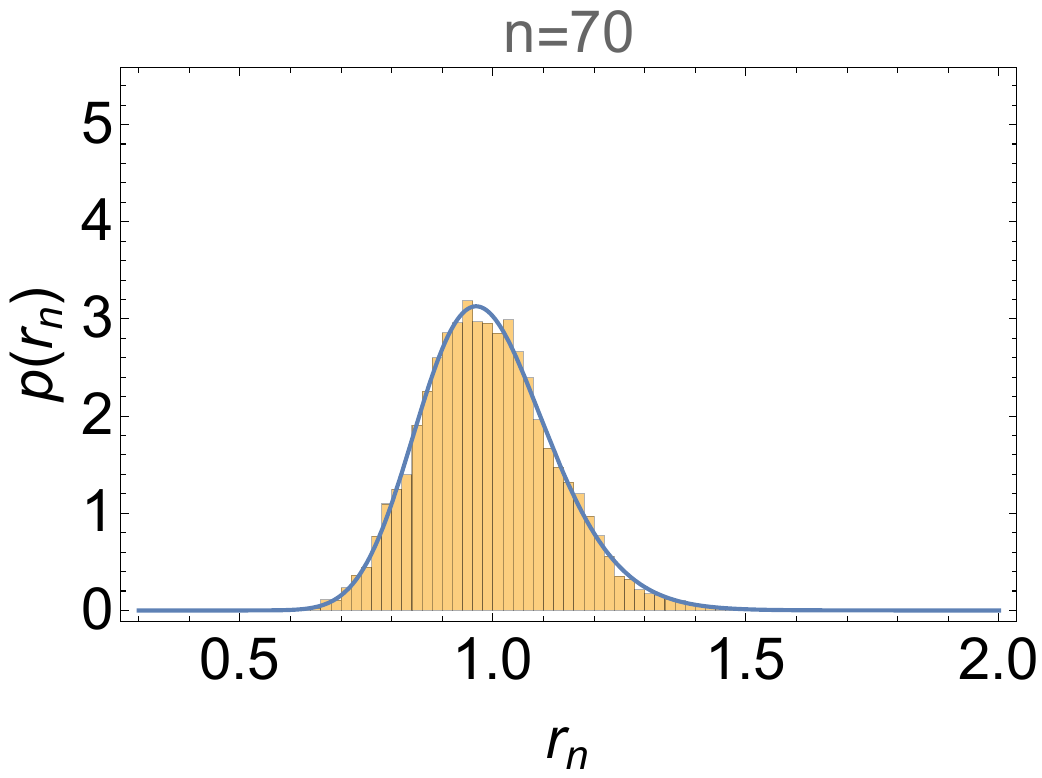} % Replace with your image file name
        %\caption*{Right: $n=50$} % Optional: subfigure caption without numbering
        %\label{fig:fig_b} % Label for right panel
    \end{minipage}
    \hfill
    \begin{minipage}[b]{0.32\textwidth}
        \centering
        \includegraphics[width=\textwidth]{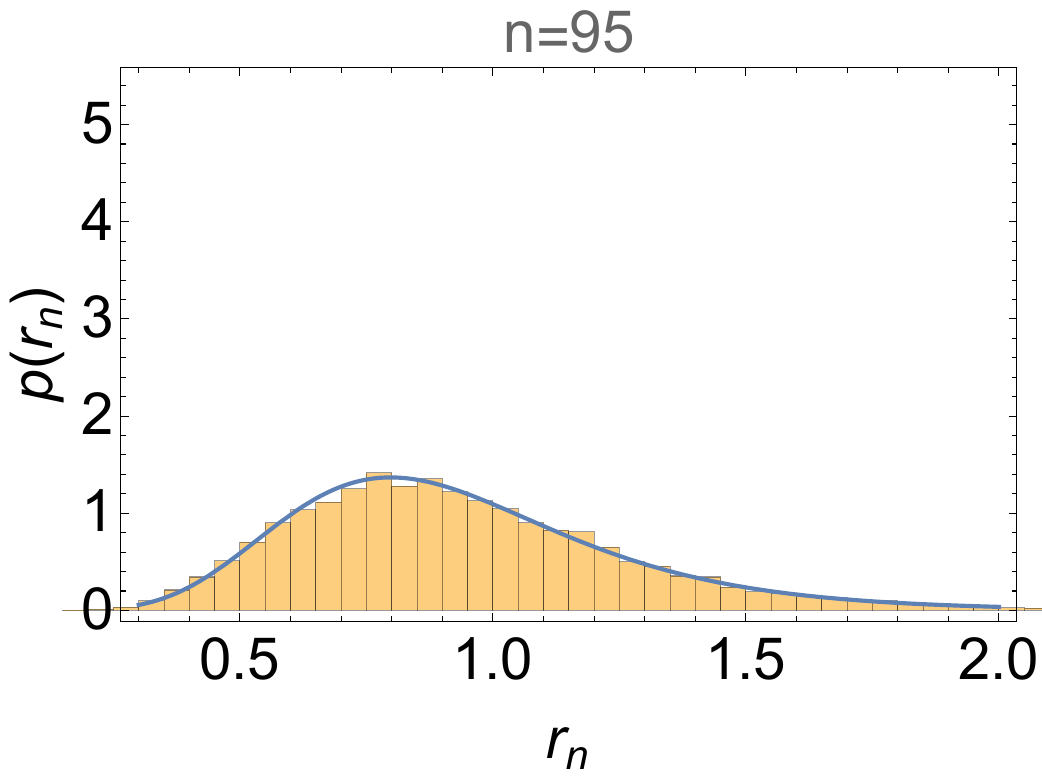} % Replace with your image file name
        %\caption*{Right: $n=50$} % Optional: subfigure caption without numbering
        %\label{fig:fig_b} % Label for right panel
    \end{minipage}
    \caption{Probability distribution function of the ratio $\mathbf{r}_n = \frac{b_{n+1}}{b_n}$ for $n=10$ (left panel) and $n=70$ (center panel) and $n=95$ (right panel). The histogram data were generated by computing $\mathbf{r}_n$ from 10,000 realizations of random matrices of size $100 \times 100$ drawn from a GUE. The initial state was chosen as $|\psi_0\rangle = (1, 0, \ldots, 0)^T$. The solid blue lines represent the analytical expression for the probability distributions of $\tbfr_n$ obtained in eq.~(\ref{eq-PDF-rn}).
    }
\label{fig:rn_distribution}
\end{figure}

We have plotted the probability distribution function for $\tbfr_n$ in eq. \eqref{eq-PDF-rn} for three different values of $n$ ($n=10$, $n=70$ and $n=95$, respectively) for a Hamiltonian drawn from GUE with $N=100$ in Fig. \ref{fig:rn_distribution} (solid blue curves). To compare these
analytical expressions with numerical results, we have used the Lanczos algorithm to first compute 
$b_n$s for an initial state $|\psi_0\rangle = (1, 0, \ldots, 0)^T$ for 10,000
different realizations of a  GUE Hamiltonian of size $100 \times 100$, and subsequently obtained
the distributions of $\tbfr_{10}$, $\tbfr_{50}$ and $\tbfr_{95}$. As can be seen, the function in eq. \eqref{eq-PDF-rn} provides a very good approximation for the numerically
obtained distributions for all three values of $n$. We also note that, for finite values of $N$, the distributions of $\tbfr_n$ have larger variances as one increases $n$ towards the end of the Krylov chain (see also the plots for the distributions of $\tbfr_n$ with $n$ in Fig. \ref{fig:four_plots}). 

%%%%%%%%%%%%%%%%%
%%%%%%%%%%%%%%%%
%%%%%%%%%%%%%%%%%

%%%%%%%%%%%%%%%%

We can  straightforwardly obtain the probability distribution function for 
$\textbf{l}_n=\log \tbfr_n$ as well by following  a similar procedure. This is given by  the expression
\begin{equation}
    p(\textbf{l}_n)= N_n ~ e^{(k-\beta)\textbf{l}_n} \Big(e^{2\textbf{l}_n}+1\Big)^{\frac{\beta}{2}-k}~. 
\end{equation}

\textbf{Distribution of Lanczos coefficients for quenched random matrices.}
Following a similar analysis as the one described in \cite{Trotter1984EigenvalueDO, dumitriu2002matrix} for $\beta=1$,
and $\beta=2$, it is possible to obtain the distribution of the Lanczos coefficients for 
the one-parameter class of the random matrices defined in \eqref{denseH} for 
the specific values of the parameter $h$ we considered in section \ref{RMT_quench}. Specifically, in the quench protocol we discussed in section \ref{GOEquench}, the pre-quench Hamiltonian is defined with $h=-1$, which is then changed to  $h=1$ after the quench, 
with both the Hamiltonian belonging to the GOE. Now, it is easy to see that each element of the $N/2\times N/2$ matrix $\tilde{B}=h B$
 are Gaussian random variables with mean zero and variance $1/N$, for both pre- and post-quench
 Hamiltonians. Therefore, when $h=\pm 1$, the distributions of the Lanczos coefficients do not
 depend on the magnitude of $h$, and for a random initial state of the form $(1, 0, \ldots, 0)^T$, are respectively, independent Gaussian distributions 
 with mean zero and variance $2/N$ for the first set of coefficients, and independent random 
 variables with distributions as in eq. \eqref{bn_distributions} for the second set. We also note that for general values of $h$ other than $\pm 1$, the analytical expressions for the distributions of the Lanczos coefficients are not known to us.

\section{Dependence of the $r$-parameter on the quench parameter $h$.}
In this appendix, we study the $r$-parameter statistics for random matrices of the form 
\begin{equation} \label{denseHA}
		H_r(h) = \left(
		\begin{array}{ccc}
			A & h B   \\
			h B^\dagger & C 
		\end{array}
		\right)~.
\end{equation}
as considered in the main text. Specifically, we study the behavior of the $r$-parameter as a function of the quench parameter $h$, for matrices belonging to the GOE and GUE when $h = \pm 1$.

The $r$-parameter serves as a tool for detecting correlations in the spectrum, thereby helping to characterize the dynamics of the system \cite{PhysRevB.75.155111,Atas_2013}. It is particularly useful because it does not require unfolding the spectrum, which would otherwise involve knowing the average density of states. Given an ordered spectrum ${ E_n }$, with $n = 1, \dots, N$, for a fixed symmetry sector of the system, we start by defining the nearest-neighbor energy spacings $s_n = E_{n+1} - E_n$. Next, we compute the ratios 
\begin{equation} 
\tilde{r}_n = \min \left( \frac{s_n}{s_{n-1}}, \frac{s_{n-1}}{s_n} \right)\,. 
\end{equation}
The $r$-parameter is then obtained as the average of these ratios: 
\begin{equation} \label{r_param_ave}
\langle \tilde{r} \rangle = \frac{1}{N-1} \sum_{n=1}^{N-1} \tilde{r}_n\,. 
\end{equation}
For an uncorrelated spectrum with level spacing statistics following a Poisson distribution, the $r$-parameter takes the value $\langle \tilde{r} \rangle_{ \text{\tiny Poisson}} \approx 0.386$. In contrast, for random matrices from the GOE, GUE, and GSE, the $r$-parameter takes the values $\langle \tilde{r} \rangle_{\text{\tiny GOE}} \approx 0.536$, $\langle \tilde{r} \rangle_{\text{\tiny GUE}} \approx 0.603$, and $\langle \tilde{r} \rangle_{\text{\tiny GSE}} \approx 0.676$, respectively. These values are computed for an ensemble average of \(3 \times 3\) random matrices, though they exhibit some dependence on \(N\). In the large-\(N\) limit, it was shown in \cite{Nishigaki:2024yjr} that $\langle \tilde{r} \rangle_{\text{\tiny GUE}} \approx 0.5997504209(1)$.

\begin{figure}[h!]
    \centering
    \begin{minipage}[b]{0.45\textwidth}
        \centering
        \includegraphics[width=\textwidth]{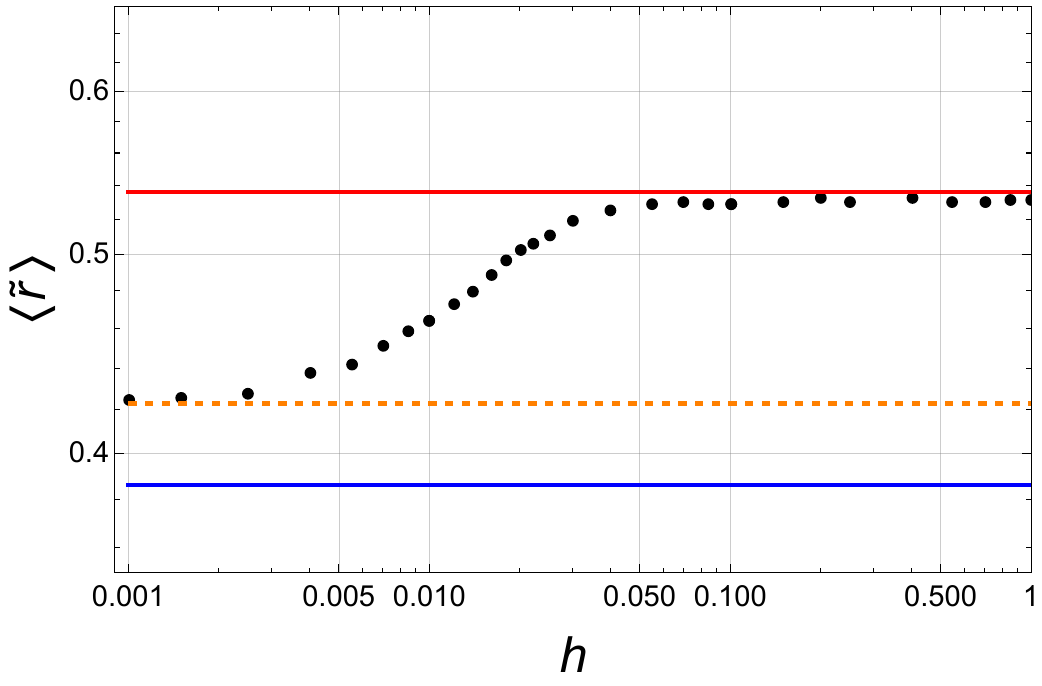}
    \end{minipage}
    \hfill
    \begin{minipage}[b]{0.45\textwidth}
        \centering
        \includegraphics[width=\textwidth]{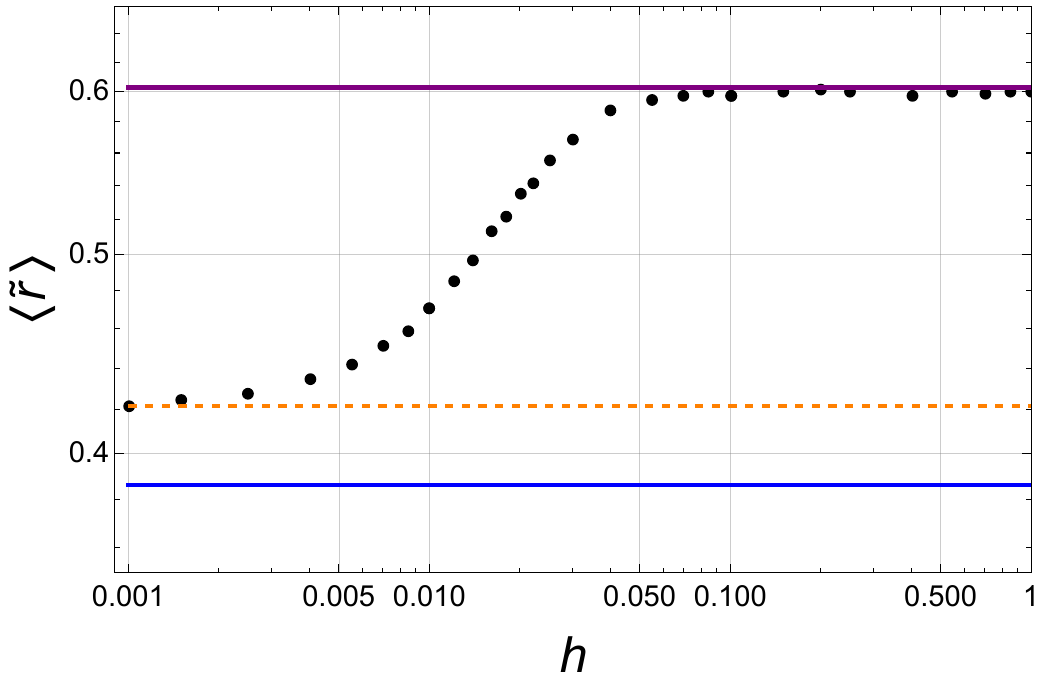} 
    \end{minipage}
    \caption{The black dots show the $r$-parameter as a function of the quench parameter $h$. The purple line represents the average value for the GUE, $\langle \tilde{r} \rangle_\text{\tiny GUE} = 0.602$, while the red line represents the average for the GOE, $\langle \tilde{r} \rangle_\text{\tiny GOE} = 0.536$. The blue line corresponds to the Poisson distribution, $\langle \tilde{r} \rangle_\text{\tiny Poisson} = 0.386$, and the orange line indicates the average value at $h=0$, $\langle \tilde{r} \rangle_{h=0} \approx 0.423$ for GOE-type quenches and $\langle \tilde{r} \rangle_{h=0} \approx 0.422$ for GUE-type quenches.  Each point is calculated from 100 realizations of $1000 \times 1000$ random matrices of the form specified in equation (\ref{denseH}), with the matrices drawn from a GOE (left panel) and GUE (right panel).}
\label{fig:rnversush}
\end{figure}

Figure~\ref{fig:rnversush} shows the $r$-parameter as a function of the quench parameter $h$ for quenches based on random matrices drawn from the GOE and GUE. At $h = 0$, the $r$-parameter attains an approximate value of $\langle \tilde{r} \rangle_{h=0} \approx 0.423$ for GOE-type quenches and $\langle \tilde{r} \rangle_{h=0} \approx 0.422$ for GUE-type quenches, matching previous results in the literature up to the third decimal place. See, for instance, Table I of \cite{Giraud:2020mmb}. This value is below the GOE and GUE averages of 0.53 and 0.60, respectively, but remains above the Poisson value of 0.386. This suggests that for $|h| < 1$, the level spacing statistics of the Hamiltonian  in \eqref{denseHA} may be approximately described by a Gaussian $\beta$-ensemble with an effective Dyson index $\beta_{\text{eff}}$ taking certain real positive values\footnote{Preliminary analysis supports this expectation, although precise fitting of the Dyson index $\beta$ remains challenging and is currently under investigation.}.

%\begin{figure}[h!]
%    \centering
%    \includegraphics[width=0.6\linewidth]{rnversusbeta.pdf}
%   \caption{The black dots represent the $r$-parameter as a function of the quench parameter $h$. The red line indicates the average for the GOE, $\langle \tilde{r} \rangle_\text{\tiny GOE} = 0.536$, while the blue line represents the Poisson distribution, $\langle \tilde{r} \rangle_\text{\tiny Poisson} = 0.386$. The orange line shows the average value when $h=0$, $\langle \tilde{r} \rangle_{h=0} = 0.423$. Each point was estimated with 100 realizations of $1000 \times 1000$ random matrices of the form specified in equation (\ref{denseH}), with block diagonal matrices derived from a GOE.}
%    \label{fig:datarn}
%\end{figure}
%%%%%%%%%%%%%%
%%%%%%%%%%%%%%%%%%%

%%%%%%%%%%%%%%%%%%%%%%%%
%%%%%%%%%%%%%%%%%%%%%%%%%
\bibliography{reference}  
\bibliographystyle{JHEP}

 \begin{comment}
	
	\end{comment}
\end{document}